\documentclass[10pt,conference,letterpaper]{IEEEtran}
\usepackage{times}
\usepackage{amsfonts}
\usepackage{graphicx}

\usepackage{url}
\usepackage{epsfig}
\usepackage{graphicx}
\usepackage{xspace}
\usepackage{subfigure}
\usepackage{float}
\usepackage{amsmath}
\usepackage{amssymb}
\usepackage{latexsym}
\usepackage{algorithm}
\usepackage[noend]{algorithmic}
\usepackage{mathrsfs}
\usepackage{wasysym}
\usepackage{color}
\usepackage{multirow}

\long\def\comment#1{}

\def\thepage{\arabic{page}}
\title{Efficient Core Maintenance in Large Dynamic Graphs}
\author{
Rong-Hua Li and Jeffrey Xu Yu
\vspace{1.6mm}\\
\fontsize{10}{10}\selectfont\itshape
The Chinese University of Hong Kong\\
\fontsize{9}{9}\selectfont\ttfamily\upshape
\{rhli, yu\}@se.cuhk.edu.hk\\
}

\newcounter{example}[section]
\renewcommand{\theexample}{\nthesection.\arabic{example}}
\newenvironment{example}{
     \refstepcounter{example}
     {\vspace{1ex} \noindent\bf  Example  \theexample:}}{
     \eop\vspace{1ex}} 

\newcounter{definition}[section]
\renewcommand{\thedefinition}{\nthesection.\arabic{definition}}
\newenvironment{definition}{
     \refstepcounter{definition}
     {\vspace{1ex} \noindent\bf  Definition  \thedefinition:}}

\newcounter{theorem}[section]
\renewcommand{\thetheorem}{\nthesection.\arabic{theorem}}
\newenvironment{theorem}{\begin{em}
        \refstepcounter{theorem}
        {\vspace{1ex} \noindent\bf  Theorem  \thetheorem:}}{
        \end{em}}\vspace{1ex} 

\newcounter{lemma}[section]
\renewcommand{\thelemma}{\nthesection.\arabic{lemma}}
\newenvironment{lemma}{\begin{em}
        \refstepcounter{lemma}
        {\vspace{1ex}\noindent\bf Lemma \thelemma:}}{
        \end{em}}

\newcounter{corollary}[section]
\renewcommand{\thecorollary}{\nthesection.\arabic{lemma}}

\newcounter{remark}[section]
\renewcommand{\theremark}{\thesection.\arabic{remark}}

\newcommand{\nthesection}{\arabic{section}}

\newcommand{\eop}{\hspace*{\fill}\mbox{$\Box$}}

\newcommand{\stitle}[1]{\vspace{1ex} \noindent{\bf #1}}

\allowdisplaybreaks
\begin{document}

\thispagestyle{plain}

\maketitle
\begin{abstract}
The $k$-core decomposition in a graph is a fundamental problem for
social network analysis. The problem of $k$-core decomposition is to
calculate the core number for every node in a graph. Previous
studies mainly focus on $k$-core decomposition in a static graph.
There exists a linear time algorithm for $k$-core decomposition in a
static graph. However, in many real-world applications such as
online social networks and the Internet, the graph typically evolves
over time. Under such applications, a key issue is to maintain the
core number of nodes given the graph changes over time. A simple
implementation is to perform the linear time algorithm to recompute
the core number for every node after the graph is updated. Such
simple implementation is expensive when the graph is very large. In
this paper, we propose a new efficient algorithm to maintain the
core number for every node in a dynamic graph. Our main result is
that only certain nodes need to update their core number given the
graph is changed by inserting/deleting an edge. We devise an
efficient algorithm to identify and recompute the core number of
such nodes. The complexity of our algorithm is independent of the
graph size. In addition, to further accelerate the algorithm, we
develop two pruning strategies by exploiting the lower and upper
bounds of the core number. Finally, we conduct extensive experiments
over both real-world and synthetic datasets, and the results
demonstrate the efficiency of the proposed algorithm.
\end{abstract}

\section{Introduction}
\label{sec:intro}In the last decade, online social network analysis
has become an important topic in both research and industry
communities due to a larger number of applications. A crucial issue
in social network analysis is to identify the cohesive subgroups of
users in a network. The cohesive subgroup denotes a subset of users
who are well-connected to one another in a network
\cite{05socialbook}. In the literature, there are a larger number of
metrics for measuring the cohesiveness of a group of users in a
social network. Examples include cliques, $n$-cliques, $n$-clans,
$k$-plexes, $k$-core, $f$-groups, $k$-trusses and so on
\cite{05trusses}.

For most of these metrics except $k$-core, the computational
complexity is typically NP-hard or at least quadratic. $k$-core, as
an exception, is a well-studied notion in graph theory and social
network analysis \cite{83kcoredef}. Through-out the paper, we will
interchangeably use graph and network. Given a graph $G$, the
$k$-core is the largest subgraph of $G$ such that all the nodes in
the $k$-core have at least degree $k$. For each node $v$ in $G$, the
core number of $v$ denotes the largest $k$-core that contains $v$.
The $k$-core decomposition in a graph $G$ is to calculate the core
number for every node in $G$. There is a linear time algorithm,
devised by Batagelj and Zaversnik \cite{03omalgkcore}, to compute
the $k$-core decomposition in a graph $G$.

Besides the analysis of cohesive subgroup, $k$-core decomposition
has been recognized as a powerful tool to analyze the structure and
function of a network, and it has many applications. For example,
the $k$-core decomposition has been applied to visualize the large
networks \cite{99kcoreviz,05nipskcoreviz}, to map, model and analyze
the topological structure of the Internet \cite{07pnaskcore,
08kcoreinternet}, to predict the function of protein in
protein-protein interaction network \cite{03proteinkcore,
03proteinfunkcore, 05proteinkcore}, to identify influential spreader
in complex networks \cite{10naturephykcore}, as well as to study
percolation on complex networks \cite{06kcorepercolation}.

\begin{figure}[t]
\begin{center}
\includegraphics[scale=0.3]{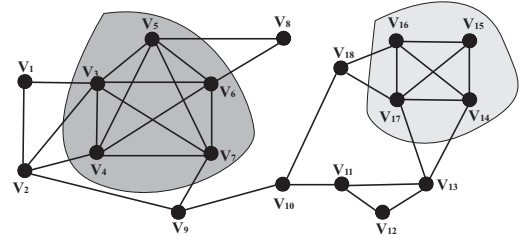}
\vspace*{-1em} \caption{An example graph.} \label{fig:kcore}
\end{center}
\vspace*{-0.8cm}
\end{figure}

From the algorithmic perspective, efficient and scalable algorithms
for $k$-core decomposition in a static graph already exist
\cite{03omalgkcore, 11icdekcorejames, 11podcdistributedcore}.
However, in many real-world applications, such as online social
network and the Internet, the network evolves over time. In such a
dynamic network, a crucial issue is to maintain the core number for
every node in a network provided the network changes over time. In a
dynamic network, it is difficult to update the core number of nodes.
The reason is as follows. An edge insertion/deletion results in the
degree of two end-nodes of the edge increase/decrease by 1. This may
lead to the updates of the core number of the end-nodes. Such
updates of the core number of the end-nodes may affect the core
number of the neighbors of the end-nodes which may need to be
updated. In other words, the update of the core number of the
end-nodes may \emph{spread} across the network. For example, in
Fig.~\ref{fig:kcore}, assume that we insert an edge $(v_8, v_{10})$
into the graph, resulting in the degree of $v_8$ and $v_{10}$
increase by 1. Suppose the core number of $v_8$ and $v_{10}$
increase by 1, then we can see that such core number update leads to
the core number of $v_{10}$'s neighbors ($v_9$, $v_{18}, v_{11}$)
that may need to be updated. And then the update of core number of
$v_{10}$'s neighbors will result in the update of core number of
$v_{10}$'s neighbors' neighbors. This update process may
\emph{spread} over the network. Therefore, it is hard to determine
which node in a network should update its core number given the
network changes.

To update the core number for every node in a dynamic graph, in
\cite{10kcoredynamic}, Miorandi and Pellegrini propose to use the
linear algorithm given in \cite{03omalgkcore} to recompute the core
number for every node in a graph. Obviously, such an algorithm is
expensive when the graph is very large. In this paper, we propose a
efficient algorithm to maintain the core number for each node in a
dynamic network. Our algorithm is based on the following key
observation. We find that only a certain number of nodes need to
update their core number when a graph is updated by
inserting/deleting an edge. Reconsider the example in
Fig.~\ref{fig:kcore}. After inserting an edge $(v_8, v_{10})$, we
can observe that only the core number of the nodes $\{v_8, v_{10},
v_{18}, v_9, v_2\}$ updates, while the core number of the remaining
nodes does not change. The key challenge is how to identify the
nodes whose core numbers need to be updated. To tackle this problem,
we propose a three-stage algorithm to update the core number of the
nodes. First, we prove that only the core number of the nodes that
are reachable from the end-nodes of the inserted/deleted edge and
their core numbers equal to the minimal core number of the end-nodes
may need to be updated. Based on this, we propose a coloring
algorithm to find such nodes whose core numbers may need to be
updated. Second, from the nodes found by the coloring algorithm, we
propose a recoloring algorithm to identify the nodes whose core
numbers definitely need to be updated. Third, we update the core
number of such nodes by a linear algorithm. The major advantage of
our algorithm is that its time complexity is independent of the
graph size, and it depends on the size of the nodes found by the
coloring algorithm. To further accelerate our algorithm, we develop
two pruning techniques to reduce the size of the nodes found by the
coloring algorithm. In addition, it is worth mentioning that our
proposed algorithm can also be used to handle a batch of edge
insertions and deletions by processing the edges one by one. Also,
the proposed technique can be applied to process node insertions and
deletions, because node insertions and node deletions can be
simulated by a sequence of edge insertions and edge deletions
respectively. Finally, we extensively evaluate our algorithm over 15
real-world datasets and 5 large synthetic datasets, and the results
demonstrate the efficiency of our algorithm. More specifically, in
real-world datasets, our algorithm reduces the average update time
over the baseline algorithm from 3.2 times to 101.8 times for
handling a single edge update. For handling a batch of edge updates,
our algorithm needs to process the edge updates one by one, while
the baseline algorithm only needs to run once for all edge updates.
In the largest synthetic dataset (5 million nodes and 25 million
edges), the results show that our algorithm is still more efficient
than the baseline algorithm when the number of edge updates is
smaller than 4700.

The rest of this paper is organized as follows. We give the problem
statement in Section~\ref{sec:problem}. We propose our basic
algorithm as well as the pruning strategies in
Section~\ref{sec:alg}. Extensive experimental studies are reported
in Section \ref{sec:experiments}, and the related work is discussed
in Section \ref{sec:rlwork}. We conclude this work in Section
\ref{sec:concl}.

\section{Preliminaries}
\label{sec:problem} Consider an undirected and unweighted graph
$G=(V, E)$, where $V$ denotes a set of nodes and $E$ denotes a set
of undirected edges between the nodes. Let $n=|V|$ and $m=|E|$ be
the number of nodes and the number of edges in $G$, respectively. A
graph $G ^\prime=(V ^\prime, E ^\prime)$ is a subgraph of $G$ if $V
^\prime \subseteq V$ and $E ^\prime \subseteq E$. We give the
definition of the $k$-core \cite{83kcoredef} as follows.

\begin{definition}
  \label{def:kcore} A $k$-core is the largest subgraph $G ^\prime$ of $G$ such
  that each node in $G ^\prime$ has at least a degree $k$.
\end{definition}
The core number of node $v$ is defined as the largest $k$-core that
contains this node. We denote the core number of node $v$ as $C_v$.
It is worth noting that the nodes with a large core number are also
in the low order core. That is to say, the cores are nested. For
example, assuming a node $v$ is in a $3$-core, then node $v$ is also
in $2$-core, $1$-core and $0$-core.

Given a graph $G$, the problem of $k$-core decomposition is to
determine the core number for every node in $G$. The following
example illustrates the concept of $k$-core composition in graph.

\begin{example}
  \label{exp:kcore}Fig.~\ref{fig:kcore} shows a graph $G$ that
  contains 18 nodes, i.e., $v_1, \cdots, v_{18}$. By
  Definition~\ref{def:kcore}, we can find that the nodes $v_3, \cdots,
  v_7$ form a $4$-core. The reason is because the induced subgraph
  by the nodes $v_3, \cdots, v_7$ is the largest subgraph in which the
  degrees of nodes are lager than or equal to 4. Similarly, the subgraph
induced by the nodes
  $v_3, \cdots, v_7, v_{14}, \cdots, v_{17}$ is a $3$-core, and the whole graph $G$ is
  a $2$-core. Here we can find that the nodes $v_3, \cdots, v_7$ are
  also in the $3$-core and $2$-core.
\end{example}

It is well known that the $k$-core decomposition in a static graph
can be calculated by a $O(n+m)$ algorithm \cite{03omalgkcore}. In
many applications such as online social networks, the graph evolves
over time. In this paper, we consider the problem of updating the
core number for every node in the graph given the graph changes over
time. In this problem, we assume that the core numbers of all the
nodes have been known before the graph is updated.  The potential
change in our problem is that either edge insertion or edge deletion
may result in the core number of a number of nodes that needs to be
updated. Previous solution for this problem \cite{10kcoredynamic} is
to perform the $O(n+m)$ core decomposition algorithm to re-compute
the core number for every node in the updated graph. Clearly, such
algorithm is expensive when the graph is very large. In the
following, we mainly focus on devising more efficient algorithm for
$k$-core decomposition in a graph given the graph is updated by an
edge insertion or deletion. Our proposed algorithm can also be used
for processing a batch of edge updates. Moreover, since node
insertions and deletions can be easily simulated as a sequence of
edge insertions and edge deletions respectively, our algorithm can
also be applied to handle node insertions and node deletions.

\section{The proposed algorithm}
\label{sec:alg} Let $N(v)$ be the set of neighbor nodes of node $v$,
$D_v$ be the degree of node $v$, i.e., $D_v = |N(v)|$. Then, we give
two important quantities associated with a node $v$ as follows.
Specifically, we define $X_v$ as the number of $v$'s neighbors whose
core numbers are greater than or equal to $C_v$, and define $Y_v$ as
the number of $v$'s neighbors whose core numbers are strictly
greater than $C_v$. Formally, for a node $v$, we have $X_v = |\{u: u
\in N(v), C_u \ge C_v\}|$ and $Y_v = |\{u: u \in N(v), C_u >
C_v\}|$. In effect, by definition, $X_v$ denotes the degree of node
$v$ in the $C_v$-core. The following lemma shows that $C_v$ is
bounded by $Y_v$ and $X_v$.

\begin{lemma}
  \label{lem:xylemma}For every node $v$ of a graph $G$, we have $Y_v \le C_v \le
  X_v \le D_v$.
\end{lemma}

\begin{myproof}
We denote the subgraph $G^\prime=(V^\prime, E^\prime)$ as the
$C_v$-core. Obviously, node $v$ is in $G^\prime$. By
Definition~\ref{def:kcore}, in $G^\prime$, node $v$ has at least
$C_v$ neighbors, and the core number of all the nodes in $G^\prime$
is at least $C_v$. In other words, the number of $v$'s neighbors
whose core numbers are larger than or equal to $C_v$ is at least
$C_v$. By definition, $X_v$ denotes such number. Therefore, we have
$C_v \le X_v$. In addition, by definition, we clearly know that $X_v
\le D_v$. For $Y_v \le C_v$, we can prove it by contradiction.
Suppose $Y_v > C_v$, then node $v$ has more than $C_v$ neighbors
whose core numbers are strictly greater than $C_v$. By
Definition~\ref{def:kcore}, the core number of node $v$ should be at
least $C_v+1$, which is a contradiction. This completes the proof.
\end{myproof}\eop

In the following, we give an example to illustrate the concepts of
$X_v$ and $Y_v$.

\begin{example}
  \label{exp:xybound}Consider the node $v_9$ in Fig.~\ref{fig:kcore}.
By definition, the core number of node $v_9$ is 2, i.e.,
$C_{v_9}=2$, and the degree of $v_9$ equals to 3, i.e., $D_{v_9}$=3.
Node $v_9$ has three neighbors ($v_2, v_7$, and $v_{10}$) whose core
number is greater than or equal to 2, and has one neighbor ($v_7$)
whose core number is strictly greater than 2. Therefore, we have
$X_{v_9}=3$ and $Y_{v_9} = 1$, which consists with
Lemma~\ref{lem:xylemma}. Similar results can be observed from other
nodes in Fig.~\ref{fig:kcore}.
\end{example}

Below, we define the notion of induced core subgraph.

\begin{definition}
  \label{def:coregraph}Given a graph $G=(V, E)$ and a node $v$, the induced core
  subgraph of node $v$, denoted as $G_{v} = (V_v, E_v)$, is a connected subgraph which consists of node $v$.
  Moreover, the core number of all the nodes in $G_v$ is equivalent
  to $C_v$.
\end{definition}

By Definition~\ref{def:coregraph}, the induced core subgraph of node
$v$ includes the nodes such that they are reachable from $v$ and
their core numbers equal to $C_v$. Based on
Definition~\ref{def:coregraph}, we define the union of two induced
core subgraphs.

\begin{definition}
  \label{def:unioncoregraph}For two nodes $u$ and $v$ and their corresponding induced core
subgraph $G_{u} = (V_u, E_u)$ and $G_{v} = (V_v, E_v)$, the union of
$G_u$ and $G_v$ is defined as $G_{u \cup v} = (V_{u \cup v}, E_{u
\cup v})$, where $V_{u \cup v} = V_v \bigcup V_u$ and $E_{u \cup v}
= \{(v_i, v_j)|(v_i, v_j) \in E, v_i \in V_{u \cup v}, v_j \in V_{u
\cup v}\}$.
\end{definition}

It is worth mentioning that the union of two induced core subgraphs
may not be connected. The following example illustrates the
definitions of induced core subgraph and union of two induced core
subgraphs.

\begin{figure}[htbp]
\begin{center}
\includegraphics[scale=0.25]{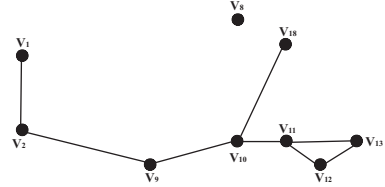}
\vspace*{-0.8em} \caption{The union of two induced core subgraphs
($G_{v_8 \cup v_{10}}$).} \label{fig:coresubgraphs}
\end{center}
\vspace*{-0.5cm}
\end{figure}

\begin{example}
  \label{exp:coresubgraph}Consider the nodes $v_8$ and $v_{10}$ in Fig.~\ref{fig:kcore}.
  By definition, the induced core subgraph of $v_8$ is a subgraph
  that only contains node $v_8$. That is to say, $V_{v_8}=\{v_8\}$ and $E_{v_8}= \emptyset$.
  The induced core subgraph of node $v_{10}$ is a subgraph that includes nodes $\{v_1, v_2, v_9, v_{10},
  v_{11},v_{12},v_{13},v_{18}\}$. In other words, $V_{v_{10}}=\{v_1, v_2, v_9, v_{10},
  v_{11},v_{12},v_{13},v_{18}\}$ and $E_{v_{10}}=\{(v_1, v_2), (v_2, v_9), (v_9, v_{10}), (v_{10}, v_{11}),
  (v_{10}, v_{18}), (v_{11}, v_{12}), \\(v_{11}, v_{13})\}$.
  The union of these two induced core subgraphs is $G_{v_8 \cup v_{10}} = (V_{v_8 \cup v_{10}},
  E_{v_8 \cup v_{10}})$, where $V_{v_8 \cup v_{10}}=\{v_1, v_2, v_8, v_9, v_{10},
  v_{11},v_{12},v_{13},v_{18}\}$ and $E_{v_8 \cup v_{10}} = E_{v_{10}}$. Fig.~\ref{fig:coresubgraphs} illustrates the union
  of two induced core subgraphs $G_{v_8 \cup v_{10}}$.
\end{example}

Based on Definition~\ref{def:coregraph} and
\ref{def:unioncoregraph}, we give a $k$-core update theorem.

\begin{theorem} (\textbf{$k$-core update theorem})
  \label{thm:kcoreupdatethm}Given a graph $G=(V, E)$ and two nodes
  $u$ and $v$.
  \begin{itemize}
    \item If $C_u > C_v$, then either insertion or deletion of an edge $(u,
  v)$ in $G$, only the core number of nodes in the
  induced core subgraph of node $v$, i.e., $G_{v}$, may need to be updated.

  \item IF $C_u < C_v$, then either insertion or deletion of an edge $(u,
  v)$ in a graph $G$, only the core number of nodes in the
  induced core subgraph of node $u$, i.e., $G_{u}$, may need to be updated.

  \item IF $C_u = C_v$, then either insertion or deletion of an edge $(u,
  v)$ in a graph $G$, only the core number of nodes in the union of two
  induced core subgraphs $G_u$ and $G_v$, i.e., $G_{u \cup v}$, may need to be updated.
  \end{itemize}
\end{theorem}

To prove Theorem~\ref{thm:kcoreupdatethm}, we first give some useful
lemmas as follows.

\begin{lemma}
  \label{lem:neighborupdate}Given a graph $G=(V, E)$ and a node $u$.
  If the core number of node $u$'s neighbors
  increases (decreases) by at most 1, then $C_u$ increases (decreases) by at most 1.
\end{lemma}

\begin{myproof}
  First, we prove the increase case by contradiction. Suppose that $C_u$ increases
   by at least 2. This implies that there are at least $C_u + 2$ neighbors
  of node $u$ whose core numbers are larger than or equal to $C_u + 2$. Since the core number of
  $u$'s neighbors increases by at most 1, the number of $u$'s neighbors
  whose core numbers are larger than or equal to $C_u + 2$ is at most $Y_u$.
  By Lemma~\ref{lem:xylemma}, we know that $Y_u \le C_u$. That is to say, the number of $u$'s neighbors
  whose core numbers are larger than or equal to $C_u + 2$ is bounded
  by $C_u$, which is a contradiction.

  Second, we prove the decrease case. If the core number of the neighbors of node $u$
  decreases by at most 1, then $u$ has at least $X_u$ neighbors whose core
  numbers are greater than or equal to $C_u - 1$. Since $X_u \ge C_u > C_u -
1$, the core number of node $u$ is at least $C_u - 1$. Therefore,
$C_u$ decreases by at most 1. This completes
  the proof.
\end{myproof}
\eop

\begin{lemma}
  \label{lem:updatecore}If we insert (delete) an edge $(u, v)$ in a graph
  $G$, the core number of any node in $G$ increases (decreases) by at most 1.
\end{lemma}

\begin{myproof}
  We focus on proving the edge insertion case, and similar arguments can be used to prove the edge deletion case.
  After inserting an edge $(u, v)$, both $D_u$ and $D_v$
  increase by 1. Recall that $X_u$ ($X_v$) denotes the degree of
  $u$ ($v$) in the $C_u$-core ($C_v$-core), which is a subgraph of $G$.
  Therefore, $X_u$ and $X_v$ increase by at most 1. By definition,
  $C_u$ ($C_v$) equals to the minimal degree of the nodes in the $C_u$-core ($C_v$-core).
  Since $X_u$ ($X_v$) increases by at most 1, the minimal degree of the nodes in the
  $C_u$-core ($C_v$-core) increases by at most 1. As a result, the core number of node
  $u$ ($v$) increases by at most 1. Such increase of $C_u$ ($C_v$) may lead to increasing the core number of the neighbors of node $u$ ($v$).
  Consider the one-hop neighbors of node $u$ ($v$). According to Lemma~\ref{lem:neighborupdate}, the core number of
all the neighbors of node $u$ ($v$) increases by at most 1. By
  recursively applying Lemma~\ref{lem:neighborupdate}, we can
  conclude that the core number of all the nodes that are reachable from $u$
  ($v$) increases by at most 1. On the other hand, the core number of the
nodes that cannot be reachable from
  $u$ ($v$) does not change. Put it all together, for any node in $G$,
  its core number increases by at most 1. This completes the
  proof.
\end{myproof}
\eop

\begin{lemma}
  \label{lem:bothinsert}Given a graph $G$ and two nodes $u$ and $v$ such that $C_u = C_v$. If we insert an edge $(u,
  v)$ in $G$, then either $C_u$ and $C_v$ increase by 1 or $C_u$ and
  $C_v$ do not change.
\end{lemma}

\begin{myproof}
  We prove it by contradiction. Without loss of generality, after
  inserting an edge $(u, v)$, we assume that $C_u$ increases by 1
  while $C_v$ does not change.
  Since $C_u$ increases by 1, node $u$ has at least $C_u + 1$
  neighbors whose core numbers are larger than or equal to $C_u + 1$.
  By Definition~\ref{def:kcore}, before inserting an edge $(u, v)$, $u$ has at most
  $C_u$ neighbors whose core numbers are larger than or equal to $C_u +
  1$. Therefore, node $v$'s core number must be $C_u + 1$, which is a
  contradiction.
\end{myproof}

\begin{lemma}
  \label{lem:waffect}Given a graph $G$ and an edge $(u, v)$.
  Suppose $G$ is updated by inserting or deleting an edge $(u, v)$.
  Then, for any node $w$ in $G$, if the core number of $w$ ($C_w$)
  needs to be changed, such change only affects the core number of nodes in
$G_w$. If $C_w$ does not change, then it does not affect
  the core number of the nodes in $G$.
\end{lemma}

\begin{myproof}
We focus on the edge insertion case, and similar proof can be used
to prove the edge deletion case. Assume that $C_w$ is changed after
inserting an edge $(u, v)$ into $G$. By Lemma~\ref{lem:updatecore},
$C_w$ increases by 1. We denote the updated $C_w$ as $\tilde C_w$,
i.e., $\tilde C_w = C_w + 1$. Obviously, the increase of $C_w$ does
not affect the core number of the nodes that cannot be reachable
from $w$. Also, we claim that the increase of $C_w$ does not affect
the core numbers of the nodes that can be reachable from $w$ and
their core numbers are less than or greater than $C_w$. First, we
consider a node $z$ that are reachable from $w$ and $C_z < C_w$.
Recall that $C_z$ equals to the minimal degree of the nodes in the
$C_z$-core. By definition, $w$ is also in the $C_z$-core (cores are
nested). The increase of $C_w$ clearly does not increase such
minimal degree. Hence, the core number of node $z$ is still $C_z$.
Second, we consider a node $z$ that is reachable from $w$ and $C_z
> C_w$. The minimal degree of the nodes in $C_z$-core is $C_z$
and $C_z \ge \tilde C_w$. Similarly, the increase of $C_w$ does not
increase such minimal degree, thereby $C_z$ will not be updated. Put
it all together, the increase of $C_w$ only affects the core number
of those nodes that are reachable from $w$ and their core numbers
equal to $C_w$, which are the nodes in $G_w$. By definition, if
$C_w$ does not change, then it will not affect the core number of
all the nodes in $G$. This completes the proof.
\end{myproof}
\eop

Armed with the above lemmas, we prove the $k$-core update theorem as
follows.

\stitle{Proof of Theorem~\ref{thm:kcoreupdatethm}: }
  For the insertion of an edge $(u, v)$, we consider three different cases: (1) $C_u > C_v$,
  (2) $C_u < C_v$, and (3) $C_u = C_v$. For $C_u > C_v$, we know
  that node $u$ is in a higher order core than node $v$. By
  Definition~\ref{def:kcore}, adding a neighbor $v$ with a small core number
  to a node $u$ does not affect $C_u$.
  By Lemma~\ref{lem:waffect}, since $C_u$ does not change, node $u$ will not affect the core number of the nodes in $G$.
  Consequently, we only need to update the core number of the nodes that
  are affected by node $v$. By Lemma~\ref{lem:waffect}, if $C_v$ changes, then
  only the core number of nodes in $G_v$ may need to be updated. If $C_v$ does not
  change, then no node's core number needs to be updated. This
  proves the case (1). Symmetrically, we can use the similar arguments to prove the case
  (2). For case (3), after inserting an edge $(u, v)$, by Lemma~\ref{lem:bothinsert}, either $C_u$ and $C_v$ increase by 1 or $C_u$ and
  $C_v$ do not change. If $C_u$ and
  $C_v$ do not change, by Lemma~\ref{lem:waffect},
  we conclude that no node's core number needs to be updated. If $C_u$
  and $C_v$ increase by 1, by Lemma~\ref{lem:waffect}, the
  core number of the nodes in $G_u$ and $G_v$ may need to be updated. That
  is to say, the core number of the nodes in $G_{u \cup v}$ may need to be
  updated.

Similarly, for the deletion of an edge $(u, v)$, we also consider
three different cases: (1) $C_u > C_v$,
  (2) $C_u < C_v$, and (3) $C_u = C_v$. The proof for the first two
  cases is very similar to the proof for the first two cases under edge insertion case, thereby we omit for brevity.
  For $C_u = C_v$, after deleting an edge $(u,
  v)$, if $C_u$ and $C_v$ do not change,
  we conclude that no node's core number needs to be updated according to Lemma~\ref{lem:waffect}. If
  $C_u$ changes, by Lemma~\ref{lem:waffect}, the core number of nodes in
  $G_u$ may need to be updated. Likewise, if
  $C_v$ changes, the core number of nodes in $G_v$ may need to be updated. To
  summarize, after removing an edge $(u,
  v)$, only the core number of the nodes in $G_{u \cup v}$ may need to be
  updated. This completes the proof.
\eop

\subsection{The basic algorithm}
\label{subsec:updatealgorithms} In this subsection, we present a
basic algorithm for core maintenance in a graph given the graph is
updated by an edge insertion or an edge deletion. Below, we describe
the detailed algorithms for edge insertion and deletion,
respectively.

\stitle{Algorithm for edge insertion: }Our main algorithm for edge
insertion consists of three steps. After inserting an edge $(u, v)$,
by the $k$-core update theorem, only the core number of nodes in the
induced core subgraph ($G_u$ or $G_v$ or $G_{u \cup v}$) may need to
be updated. Therefore, the first step of our main algorithm is to
identify the nodes in the induced core subgraph. Let $V_c$ be the
set of nodes found in the first step. Then, the second step of our
algorithm is to determine those nodes in $V_c$ whose core numbers
definitely need to be updated. Finally, the third step of our
algorithm is to update the core number of such nodes.

Our main algorithm for edge insertion, called \textbf{Insertion}, is
outlined in Algorithm~\ref{alg:insertion}.
Algorithm~\ref{alg:insertion} includes three sub-algorithms, namely
\textbf{Color}, \textbf{RecolorInsert}, and \textbf{UpdateInsert},
which corresponds the first, the second, and the third step of our
main algorithm, respectively. In particular, \textbf{Color} is used
to color the nodes in $V_c$ with a color 1, \textbf{RecolorInsert}
is applied to recolor the nodes in $V_c$ whose core numbers are
definitely unchanged with a color 0, and \textbf{UpdateInsert} is
used to update the core number of the nodes in $V_c$ with a color 1.
The detailed description of Algorithm~\ref{alg:insertion} is as
follows. First, Algorithm~\ref{alg:insertion} assigns a color 0 for
every node in $G$ (line~2 in Algorithm~\ref{alg:insertion}) and
initializes $V_c$ by an empty set (line~3 in
Algorithm~\ref{alg:insertion}). Second, the algorithm updates the
core number of the nodes under three different cases, i.e., $C_u
> C_v$, $C_u < C_v$, and $C_u = C_v$. Specifically, under the first
case ($C_u > C_v$), the algorithm first invokes \textbf{Color}($G$,
$v$, $c$) to find the nodes in $G_v$ (line~6 in
Algorithm~\ref{alg:insertion}), because only the core number of the
nodes in $G_v$ may need to be updated. After this process, all the
nodes in $G_v$ are recorded in $V_c$ and all of them are colored by
1. Then, the algorithm invokes \textbf{RecolorInsert}($G$, $c$) to
identify the nodes whose core numbers are definitely unchanged
(line~7 in Algorithm~\ref{alg:insertion}). After this step, all of
such nodes in $V_c$ are recolored by 0. Finally, the algorithm
invokes \textbf{UpdateInsert}($G$, $c$) to update the core number of
the nodes in $V_c$ with color 1 (line~8 in
Algorithm~\ref{alg:insertion}). Similar process can be used for
other two cases (line~9-13 in Algorithm~\ref{alg:insertion}). Note
that for the case $C_u = C_v$, we can invoke \textbf{Color}($G$,
$u$, $c$) to find the nodes in $G_{u \cup v}$, because $u$ can reach
$v$ after inserting an edge $(u, v)$. Below, we describe the details
of our sub-algorithms, \textbf{Color}, \textbf{RecolorInsert}, and
\textbf{UpdateInsert}, respectively.

\begin{algorithm}[t]
\caption{ \textbf{Insertion}($G$, $u$, $v$)} \label{alg:insertion}
{\small
\begin{tabbing}
    {\bf\ Input}: \hspace{0.3cm}\= Graph $G=(V, E)$ and an edge $(u, v)$ \\
{\bf\ Output}: the updated core number of the nodes\>
\end{tabbing}
\begin{algorithmic}[1]
\STATE Initialize visited($w$) $\leftarrow 0$ for all node $w \in
V$; \STATE Initialize color($w$) $\leftarrow 0$ for all node $w \in
V$; \STATE $V_c \leftarrow \emptyset$; \IF {$C_u$  $> C_v$}
    \STATE $c \leftarrow C_v$;
    \STATE \textbf{Color}($G$, $v$, $c$);
    \STATE \textbf{RecolorInsert}($G$, $c$);
    \STATE \textbf{UpdateInsert}($G$, $c$);
\ELSE
    \STATE $c \leftarrow C_u$;
    \STATE \textbf{Color}($G$, $u$, $c$);
    \STATE \textbf{RecolorInsert}($G$, $c$);
    \STATE \textbf{UpdateInsert}($G$, $c$);
\ENDIF
\end{algorithmic}
}
\end{algorithm}

\begin{algorithm}[t]
\caption{void \textbf{Color}($G$, $u$, $c$)} \label{alg:color}
{\small
\begin{algorithmic}[1]
\STATE visited($u$) $\leftarrow 1$; \IF {color($u$) = 0}
    \STATE $V_c \leftarrow V_c \cup \{u\}$;
    \STATE color($u$) = 1;
\ENDIF
\FOR {each node $w \in N(u)$}
    \IF {visited($w$) $= 0$ and $C_w = c$}
        \STATE \textbf{Color}($G$, $w$, $c$);
    \ENDIF
\ENDFOR
\end{algorithmic}
}
\end{algorithm}

\begin{algorithm}[t]
\caption{void \textbf{RecolorInsert}($G$, $c$)}
\label{alg:recolorinsert} {\small
\begin{algorithmic}[1]
\STATE flag $\leftarrow 0$; \FOR {each node $u \in V_c$}
    \IF {color($u$) = 1}
        \STATE $X_u \leftarrow 0$;
        \FOR{each node $w \in N(u)$}
            \IF{(color($w$) = 1) or ($C_w > c$)}
                \STATE $X_u \leftarrow X_u + 1$;
            \ENDIF
        \ENDFOR
        \IF{$X_u \le c$}
            \STATE color($u$) $\leftarrow 0$;
            \STATE flag $\leftarrow 1$;
        \ENDIF
    \ENDIF
\ENDFOR \IF{flag = 1}
    \STATE \textbf{RecolorInsert}($G$, $c$);
\ENDIF
\end{algorithmic}
}
\end{algorithm}

\begin{algorithm}[t]
\caption{void \textbf{UpdateInsert}($G$, $c$)}
\label{alg:updateinsert} {\small
\begin{algorithmic}[1]
\FOR {each node $w \in V_c$}
    \IF {color($w$) $= 1$}
        \STATE $C_w \leftarrow c + 1$;
    \ENDIF
\ENDFOR
\end{algorithmic}
}
\end{algorithm}

Recall that after inserting an edge $(u, v)$, by the $k$-core update
theorem, we have three cases that need to be considered, i.e., $C_u
< C_v$, $C_u > C_v$, and $C_u = C_v$. To simplify our description,
we mainly focus on describing our sub-algorithms under the case $C_u
= C_v$, and similar description can be used for other cases. Suppose
that node $u$ and $v$ have core number $C_u = C_v = c$. In this
case, we have $V_c = V_{u \cup v}$. By
Definition~\ref{def:unioncoregraph}, finding the nodes in $V_{u \cup
v}$ can be done by a Depth-First-Search (DFS) algorithm.
\textbf{Color} depicted in Algorithm~\ref{alg:color} is indeed such
a DFS algorithm. In particular, \textbf{Color} will assign a color 1
to every node in $V_c$. At the beginning, $V_c$ is initialized by an
empty set and all the nodes are associated with a color 0. The
algorithm recursively finds the nodes that are reachable from $u$
and have core number $c$ (line~6-7 in Algorithm~\ref{alg:color}).
When the algorithm visits such a node, if its color is 0, then the
algorithm colors it by 1 and adds it into the set $V_c$ (line~3-4 in
Algorithm~\ref{alg:color}). To find all the nodes in $V_{u \cup v}$,
we can invoke \textbf{Color}($G$, $u$, $c$). Recall that after
inserting edge $(u, v)$, the nodes that are reachable from $v$ can
also be
found by \textbf{Color}($G$, $u$, $c$). 

\textbf{RecolorInsert} described in
Algorithm~\ref{alg:recolorinsert} is used to identify the nodes in
$V_c$ whose core numbers are definitely unchanged. Specifically,
Algorithm~\ref{alg:recolorinsert} recursively recolors the nodes
whose core numbers do not change by a color 0. The recursion is
terminated until no node needs to be recolored. In each recursion,
the algorithm re-computes $X_u$ for each node $u$ in $V_c$. Here the
recomputed $X_u$ equals to the sum of the number of neighbors of
node $u$ whose core numbers are larger than $c$ and the number of
neighbors of node $u$ with color 1 (line~4-7 in
Algorithm~\ref{alg:recolorinsert}). For a node $u$, if the current
$X_u$ is smaller than or equal to $c$, then the algorithm recolors
it by 0 (line~8-10 in Algorithm~\ref{alg:recolorinsert}).

The rationale of Algorithm~\ref{alg:recolorinsert} is as follows.
First, Algorithm~\ref{alg:recolorinsert} assumes that the core
numbers of all the nodes in $V_c$ need to be updated. Then, for each
node $w$ in $V_c$, the algorithm recomputes $X_w$. Initially, since
all the neighbors of $w$ whose core numbers equal to $c$ are colored
by 1, $X_w$ is indeed the same value as our previous definition. If
$X_w \le c$, then $w$ at most $c$ neighbors whose core numbers are
larger than $c$ after inserting an edge $(u, v)$. As a result, $C_w$
cannot be updated and the algorithm recolors it by 0. This
recoloring process may affect the color of $w$'s neighbors. The
reason is because, before recoloring $w$, $w$ may contribute to
calculate $X_z$, where $z$ is a neighbor of $w$. Consequently, the
algorithm needs to recursively recolor the nodes in $V_c$. Note that
Algorithm~\ref{alg:recolorinsert} is recursively invoked at most
$|V_c|+1$ times, because the algorithm at least recolors one node at
a recursion in the worse case. The following theorem shows that
after Algorithm~\ref{alg:recolorinsert} terminates, a node with a
color 1 is a sufficient and necessary condition for updating its
core number.

\begin{theorem}
  \label{thm:correctrecolor}
Under the case of insertion of an edge $(u, v)$, the core number of
a node needs to be updated if and only if its color is 1 after
Algorithm~\ref{alg:recolorinsert} terminates.
\end{theorem}

\begin{myproof}
First, we prove that if the core number of a node $w$ needs to be
updated, then its color is 1 after Algorithm~\ref{alg:recolorinsert}
terminates. We focus on the case of $C_u = C_v = c$, similar proof
can be used to prove the other two cases. By our assumption and
Lemma~\ref{lem:waffect}, we have $w \in V_c$, where $V_c = V_{u \cup
v}$. Then, by Lemma~\ref{lem:updatecore}, after inserting an edge
$(u, v)$, the core number of the nodes in $V_c$ increases by at most
1. Therefore, if $C_w$ needs to be updated, then the updated core
number of $w$ must be $c + 1$. That is to say, node $w$ must have $c
+ 1$ neighbors whose core numbers are larger than or equal to $c +
1$. Now assume that the color of node $w$ is 0. This means that $X_w
\le c$ when Algorithm~\ref{alg:recolorinsert} terminates. Recall
that $X_w$ denotes to the sum of the number of neighbors whose core
numbers are larger than $c$ and the number of neighbors whose color
is 1. This implies that node $w$ has at most $c$ neighbors whose
core numbers are larger than $c$, which is a contradiction.

Second, we prove that if a node has a color 1 after
Algorithm~\ref{alg:recolorinsert} terminates, then the core number
of this node must be updated. We consider the induced subgraph by
the nodes with color 1 after Algorithm~\ref{alg:recolorinsert}
terminates and the nodes whose core numbers are greater than $c$.
Consider a node $w$ in such an induced subgraph. Clearly, if $w$ has
a color 1, then it has $X_w > c$ neighbors. And if $w$ has a color
0, then its core number $C_w$ is larger than $c$. By
Definition~\ref{def:kcore}, the induced subgraph belongs to the $(c
+ 1)$-core. Therefore, the core number of a node $w$ with color 1 is
at least $c + 1$. By Lemma~\ref{lem:updatecore}, after inserting an
edge $(u, v)$, the core number of any nodes in graph $G$ increases
by at most 1. Consequently, the core number of the nodes with color
1 increases by 1. This completes the proof.
\end{myproof}
\eop

\textbf{UpdateInsert} outlined in Algorithm~\ref{alg:updateinsert}
increases the core numbers of the nodes in $V_c$ with label 1 to
$c+1$, because only the core numbers of those nodes need to increase
by 1 after the coloring and recoloring processes. The correctness of
our algorithm for edge insertion can be guaranteed by
Theorem~\ref{thm:kcoreupdatethm} and
Theorem~\ref{thm:correctrecolor}. The following example explains how
the \textbf{Insertion} algorithm works.

\begin{figure}[t]
\begin{center}
\includegraphics[scale = 0.3]{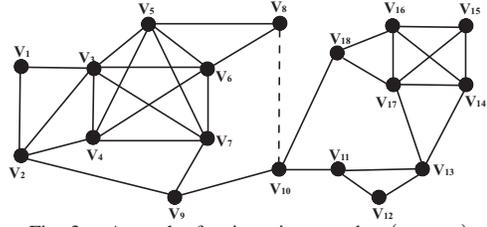}
\vspace*{-1em} \caption{A graph after inserting an edge $(v_8,
v_{10})$.} \label{fig:kcoreupdate}
\end{center}
\vspace*{-0.8cm}
\end{figure}

\begin{example}
  \label{exp:insertionalg}Let us consider the same graph given in
  Fig.~\ref{fig:kcore}. Assume that we insert an edge $(v_8,
  v_{10})$, which results in a graph given in Fig.~\ref{fig:kcoreupdate}. In Fig.~\ref{fig:kcoreupdate}, the
  dashed line denotes the inserted edge. Since $C_{v_8}=C_{v_{10}}= c = 2$,
  the \textbf{Insertion} algorithm first invokes \textbf{Color}($G$, $v_8$,
  $2$). After this process, we can get that $V_c=\{v_8, v_{10}, v_9, v_2, v_1, v_{18}, v_{11}, v_{12},
  v_{13}\}$. And all the nodes in $V_c$ are colored by 1 and the other
  nodes are colored by 0. Then, the algorithm invokes the \textbf{RecolorInsert}($G$,
  $2$) algorithm. To simplify our description, we assume that the
  node visiting-order in $V_c$ is their DFS visiting-order.
  At the first recursion, we can find that $X_{v_1} =
  2$, thereby it is recolored by 0. Also, the node $V_{12}$ is
  recolored by 0, because $X_{v_{12}}=2$. At the second recursion, we can
  find that the nodes $v_{11}$ and $v_{13}$ are recolored by 0. At
  the third recursion, no node needs to be recolored,
  the algorithm therefore terminates. After invoking the \textbf{RecolorInsert}($G$,
  $2$) algorithm, the nodes $\{v_8, v_{10}, v_9, v_2, v_{18}\}$ are
  colored by 1, thereby their core numbers must increase to 3 by Theorem~\ref{thm:correctrecolor}.
  Finally, the \textbf{Insertion} algorithm invokes
  the \textbf{UpdateInsert}($G$, $2$) algorithm to update the core
  number of such nodes. As a consequence, the core number of the nodes $\{v_8, v_{10}, v_9, v_2, v_{18}\}$
  is increased to 3.
\end{example}

We analyze the time complexity of the \textbf{Insertion} algorithm
as follows. First, the \textbf{Color} algorithm takes
$O(\sum\nolimits_{u \in V_c } {D_u } )$ time complexity. Second, the
\textbf{RecolorInsert} algorithm takes $O(|V_c|\sum\nolimits_{u \in
V_c } {D_u } )$ time complexity in the worse case, as the algorithm
is recursively invoked at most $O(|V_c|)$ times and each recursion
takes $O(\sum\nolimits_{u \in V_c } {D_u } )$ time complexity.
Finally, the \textbf{UpdateInsert} algorithm takes $O(|V_c|)$ time
complexity. Put it all together, the time complexity of the
\textbf{Insertion} algorithm is $O(|V_c|\sum\nolimits_{u \in V_c }
{D_u } )$ in the worse case, which is independent of the graph size.
However, in practice, the algorithm is more efficient than such
worse-case time complexity. The reason could be of twofold. On the
one hand, $|V_c|$ typically not very large w.r.t.\ the number of
nodes of the graph. On the other hand, very often, the
\textbf{RecolorInsert} algorithm terminates very fast.

\stitle{Algorithm for edge deletion: }The main algorithm for edge
deletion, namely \textbf{Deletion}, is outlined in
Algorithm~\ref{alg:deletion}. Similar to the edge insertion case,
\textbf{Deletion} also includes three sub-algorithms:
\textbf{Color}, \textbf{RecolorDelete}, and \textbf{UpdateDelete}.
Here \textbf{Color} is used to find the nodes in the induced core
subgraph, \textbf{RecolorDelete} is utilized to identify the nodes
whose core numbers need to be updated, and \textbf{UpdateDelete} is
applied to update the core numbers of the nodes identified by
\textbf{RecolorDelete}. The detailed description of
\textbf{Deletion} is given as follows.

Similarly, let $V_c$ be a set of nodes whose core numbers may need
to be updated. First, Algorithm~\ref{alg:deletion} initializes the
color of all the nodes to 0 and $V_c$ to an empty set. Likewise,
under the edge deletion case, we also have to consider three cases.
That is, $C_u > C_v$, $C_u < C_v$, and $C_u = C_v$. If $C_u > C_v$,
only the core number of the nodes in $G_v$ may need to be updated.
Under this case, the algorithm invokes \textbf{Color}($G$, $v$, $c$)
to find the nodes in $V_c$ (line~6 in Algorithm~\ref{alg:deletion}).
Then, the algorithm invokes \textbf{RecolorDelete}($G$, $c$) to
identify the nodes whose core numbers need to be changed (line~7 in
Algorithm~\ref{alg:deletion}). Finally, the algorithm invokes
\textbf{UpdateDelete}($G$, $c$) to update the core number of such
nodes (line~8 in Algorithm~\ref{alg:deletion}). Similar process can
be used for the $C_u < C_v$ case. For the $C_u = C_v$ case, the
algorithm first invokes \textbf{Color}($G$, $u$, $c$) to find the
nodes in $G_u$ (line~16 in Algorithm~\ref{alg:deletion}). Then, the
algorithm has to handle two different cases. First, if $u$ can reach
$v$, then the coloring algorithm can also find the nodes in $G_v$
(in this case, $v$'s color is 1, line~22 in
Algorithm~\ref{alg:deletion}). Second, if $u$ cannot reach $v$
($v$'s color is 0), then the algorithm invokes \textbf{Color}($G$,
$v$, $c$) to find the nodes in $G_v$ (line~19 in
Algorithm~\ref{alg:deletion}). After this process, all the node in
$G_{u \cup v}$ are recorded in $V_c$. Then, we can invoke
 \textbf{RecolorDelete}($G$, $c$) and \textbf{UpdateDelete}($G$,
 $c$) algorithms to update the core number of the nodes in $V_c$.
Below, we give the detailed descriptions of \textbf{RecolorDelete}
and \textbf{UpdateDelete} respectively.

\begin{algorithm}[t]
\caption{ \textbf{Deletion}($G$, $u$, $v$)} \label{alg:deletion}
{\small
\begin{tabbing}
    {\bf\ Input}: \hspace{0.3cm}\= Graph $G=(V, E)$ and an edge $(u, v)$ \\
{\bf\ Output}: the updated core number of the nodes\>
\end{tabbing}
\begin{algorithmic}[1]
\STATE Initialize visited($w$) $\leftarrow 0$ for all node $w \in
V$; \STATE Initialize color($w$) $\leftarrow 0$ for all node $w \in
V$; \STATE $V_c \leftarrow \emptyset$; \IF {$C_u$  $> C_v$}
    \STATE $c \leftarrow C_v$;
    \STATE \textbf{Color}($G$, $v$, $c$);
    \STATE \textbf{RecolorDelete}($G$, $c$);
    \STATE \textbf{UpdateDelete}($G$, $c$);
\ENDIF \IF {$C_u$  $< C_v$}
    \STATE $c \leftarrow C_u$;
    \STATE \textbf{Color}($G$, $u$, $c$);
    \STATE \textbf{RecolorDelete}($G$, $c$);
    \STATE \textbf{UpdateDelete}($G$, $c$);
\ENDIF \IF {$C_u$  $= C_v$}
    \STATE $c \leftarrow C_u$;
    \STATE \textbf{Color}($G$, $u$, $c$);
    \IF{color($v$) $= 0$}
        \STATE Initialize visited($w$) $\leftarrow 0$ for all node $w \in V$;
        \STATE \textbf{Color}($G$, $v$, $c$);
        \STATE \textbf{RecolorDelete}($G$, $c$);
        \STATE \textbf{UpdateDelete}($G$, $c$);
    \ELSE
        \STATE \textbf{RecolorDelete}($G$, $c$);
        \STATE \textbf{UpdateDelete}($G$, $c$);
    \ENDIF
\ENDIF
\end{algorithmic}
}
\end{algorithm}

\begin{algorithm}[t]
\caption{void \textbf{RecolorDelete}($G$, $c$)}
\label{alg:recolordelete} {\small
\begin{algorithmic}[1]
\STATE flag $\leftarrow 0$; \FOR {each node $u \in V_c$}
    \IF {color($u$) = 1}
        \STATE $X_u \leftarrow 0$;
        \FOR{each node $w \in N(u)$}
            \IF{(color($w$) = 1) or ($C_w > c$)}
                \STATE $X_u \leftarrow X_u + 1$;
            \ENDIF
        \ENDFOR
        \IF{$X_u < c$}
            \STATE color($u$) $\leftarrow 0$;
            \STATE flag $\leftarrow 1$;
        \ENDIF
    \ENDIF
\ENDFOR \IF{flag = 1}
    \STATE \textbf{RecolorDelete}($G$, $c$);
\ENDIF
\end{algorithmic}
}
\end{algorithm}

\begin{algorithm}[t]
\caption{void \textbf{UpdateDelete}($G$, $c$)}
\label{alg:updatedelete} {\small
\begin{algorithmic}[1]
\FOR {each node $w \in V_c$}
    \IF {color($w$) $= 0$}
        \STATE $C_w \leftarrow c - 1$;
    \ENDIF
\ENDFOR
\end{algorithmic}
}
\end{algorithm}

Similar to the edge insertion case, after invoking \textbf{Color},
the nodes whose core numbers may need to be updated are recorded in
a set $V_c$, and also all of them are colored by 1. After obtaining
the set $V_c$, \textbf{RecolorDelete} described in
Algorithm~\ref{alg:recolordelete} is used to determine the nodes
whose core numbers must be updated. In particular,
\textbf{RecolorDelete} recursively recolors the nodes whose core
numbers need to be updated by 0. In each recursion, the algorithm
calculates $X_w$ for every node $w$ in $V_c$. Here $X_w$ denotes the
sum of the number of $w$'s neighbors whose color is 1 and the number
of $w$'s neighbors whose core numbers are larger than $c$, where $c
= \min \{C_u, C_v\}$. For a node $w \in V_c$, if $X_w < c$, then the
algorithm colors $w$ by 0. The algorithm terminates if no node needs
to be recolored. Clearly, the algorithm is invoked at most $|V_c|$
times. The following theorem shows that a node in $V_c$ with a color
0 after Algorithm~\ref{alg:recolordelete} terminates is a sufficient
and necessary condition for updating its core number.

\begin{theorem}
  \label{thm:recolordelete}Under the case of deletion of an edge $(u,
  v)$, a node in $V_c$ whose core number needs to update if and only if its
  color is 0 after Algorithm~\ref{alg:recolordelete} terminates.
\end{theorem}

\begin{myproof}
First, we prove that if a node $w$ in $V_c$ whose core number needs
to be updated, then its color is 0 after
Algorithm~\ref{alg:recolordelete} terminates. By our assumption and
Lemma~\ref{lem:updatecore}, after deleting an edge $(u, v)$, $C_w$
decreases by 1. This means that $C_w$ decreases to $c - 1$. That is
to say, $w$ has $c - 1$ neighbors whose core numbers are larger than
or equal to $c - 1$. Suppose that the color of $w$ is 1 after the
algorithm terminates. This implies that $X_w \ge c$. Recall that
$X_w$ denotes the sum of the number of $w$'s neighbors whose core
numbers are lager than $c$ and the number of $w$'s neighbors whose
color is 1. Note that a node with color 1 suggests that its core
number equals to $c$. As a result, $w$ has at least $c$ neighbors
whose core numbers are larger than or equal to $c$, which is a
contradiction.

Second, we prove that if a node $w$ in $V_c$ is recolored by 0 after
Algorithm~\ref{alg:recolordelete} terminates, then $C_w$ must be
updated. After deleting an edge $(u, v)$, we construct an induced
subgraph, which is denoted as $\tilde G = (\tilde V, \tilde E)$, by
the nodes in $V_c$ and the nodes whose core numbers are larger than
$c$. Note that the core number of the nodes in $V \backslash \tilde
V$ is smaller than $c$. Therefore, they do not affect the core
number of the nodes in $\tilde V$. If a node $w \in V_c$ with a
color 0 after Algorithm~\ref{alg:recolordelete} terminates, then
$X_w < c$. This suggests that the node $w$ in $\tilde G$ has at most
$c - 1$ neighbors. By Definition~\ref{def:kcore}, $\tilde G$ at most
belongs to the $(c - 1)$-core. By Lemma~\ref{lem:updatecore}, the
core number of any nodes in $G$ decreases by at most 1 after
deleting an edge. Therefore, the core number of the nodes with color
0 decreases by 1. This completes the proof.
\end{myproof}
\eop

 \textbf{UpdateDelete} which is depicted
in Algorithm~\ref{alg:updatedelete} is used to update the core
number of the nodes in $V_c$ with color 0 to $c-1$, because only the
core numbers of those nodes need to decrease by 1 after the coloring
and recoloring steps. The correctness of \textbf{Deletion} can be
guaranteed by Theorem~\ref{thm:kcoreupdatethm} and
Theorem~\ref{thm:recolordelete}. By a similar analysis as the edge
insertion case, the time complexity of \textbf{Deletion}($G$, $u$,
$v$) is $O(|V_c|\sum\nolimits_{u \in V_c } {D_u } )$. The following
example explains how \textbf{Deletion} works.

\begin{example}
  \label{exp:deletion}Let's consider the graph depicted in
  Fig.~\ref{fig:kcoreupdate}. Suppose that we delete the
  edge $(v_8, v_{10})$. Since $C_{v_8} = C_{v_{10}} = c = 3$, the \textbf{Deletion} algorithm
  first invokes \textbf{Color}($G$, $v_8$, $3$), which results in $V_c =
  \{v_8\}$. Clearly, the color of $v_{10}$ is 0 after this process
  ends. Hence, the algorithm invokes \textbf{Color}($G$, $v_{10}$,
  $3$), which leads to $V_c = \{v_8, v_{10}, v_9, v_2, v_{18}\}$.
  After this process, all the nodes in $V_c$ are colored by 1 and
  other nodes are colored by 0. Then, the algorithm invokes \textbf{RecolorDelete}($G$,
  $3$). At the first recursion, since $X_{v_8} = 2$, $v_8$ is
  recolored by 0. Similarly, $v_{10}, v_9, v_2$, and $v_{18}$ will
  be recolored by 0 at the first recursion. At the second recursion,
  the algorithm terminates because no node needs
  to be recolored. Therefore, all the nodes in $V_c$ are recolored by 0.
  Finally, the algorithm invokes \textbf{UpdateDelete}($G$,
 $c$) to decrease the core number of all the nodes in $V_c$ to 2.
\end{example}

\subsection{Pruning strategies}
\label{subsec:prun}As analysis in the previous subsection, the time
complexity of our \textbf{Insertion} and \textbf{Deletion}
algorithms depend on the size of $V_c$. In this subsection, to
further accelerate our algorithms, we devise two pruning techniques,
namely $X$-pruning and $Y$-pruning, to remove the nodes in $V_c$
whose core numbers are definitely unchanged given the graph is
updated. 

\stitle{$X$-pruning: }By Lemma~\ref{lem:xylemma}, for a node $w$,
$X_w$ is an upper bound of $C_w$. Here we make use of such upper
bound to develop pruning technique. We refer to it as $X$-pruning.
Below, we discuss the $X$-pruning technique over the edge insertion
and edge deletion cases, respectively.

First, we consider the insertion case. Assume that we insert an edge
$(u_0, v_0)$. Also, we need to consider three cases, $C_{u_0} >
C_{v_0}$, $C_{u_0} < C_{v_0}$, $C_{u_0} = C_{v_0}$. Below, we mainly
focus on describing the $X$-pruning rule under the case of
$C_{u_0}=C_{v_0}$, and similar descriptions can be used for other
two cases. For a node $w$ in $V_c$, after inserting an edge $(u_0,
v_0)$, if $X_w$ equals to $c$, then $C_w$ cannot increases to $c+1$.
As a result, we can safely prune $w$. For example, consider an graph
in Fig.~\ref{fig:kcoreupdate}. Assume that we insert an edge $(v_8,
v_{10})$. Then, for the node $v_1$, we have $X_{v_1}=2$. Clearly,
$C_{v_1}$ cannot increase to 3, thereby we can prune $v_1$.

\begin{figure}[t]
\begin{center}
\includegraphics[scale=0.45]{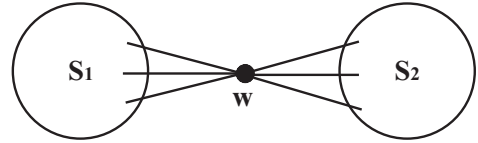}
\vspace*{-0.8em} \caption{A toy induced core subgraph.}
\label{fig:toygraph} \vspace*{-0.5cm}
\end{center}
\end{figure}

In effect, after removing $w$, for the nodes that cannot be
reachable from $u_0$ and $v_0$ in the induced core subgraph can also
be pruned. Let us consider a toy induced subgraph shown in
Fig.~\ref{fig:toygraph}. Suppose that the induced subgraph can be
partitioned into three parts, $S_1$, $w$, and $S_2$. Further, we
assume that both $u_0$ and $v_0$ are in $S_1$, and $X_w=c$. Recall
that after inserting an edge $(u_0, v_0)$, if $X_w =c$, then $C_w$
is unchanged. By Lemma~\ref{lem:waffect}, $w$ will not affect the
core numbers of the nodes in $S_2$. As a consequence, the core
numbers of the nodes in $S_2$ cannot be increased, and we can safely
prune all the nodes in $S_2$. More formally, we give a pruning
theorem as follows.

\begin{theorem}
  \label{thm:xpruninsert}Given a graph $G$ and an edge $(u_0, v_0)$. After inserting an edge $(u_0, v_0)$ in $G$,
  for a node $w \in V_c$ and $X_w < c + 1$, we have the
  following pruning rules.
\begin{itemize}
  \item If $C_{u_0} > C_{v_0}$ (i.e., $V_c = V_{{v_0}}$), then for
  any node $u \in V_c$ that every path from $v_0$ to $u$ in $G_{v_0}$ must go
  through $w$ can be pruned.

  \item If $C_{u_0} < C_{v_0}$ (i.e., $V_c = V_{{u_0}}$), then for
  any node $u \in V_c$ that every path from $u_0$ to $u$ in $G_{u_0}$ must go
  through $w$ can be pruned.

  \item If $C_{u_0} = C_{v_0}$ (i.e., $V_c = V_{{u_0 \cup v_0}}$),
  then for any node $u \in V_c$ that every path either from $u_0$ to $u$ or from $v_0$
  to $u$ in $G_{u_0 \cup v_0}$ must go through $w$ can be pruned.
\end{itemize}
\end{theorem}

\begin{myproof}
We prove this theorem under the case $C_{u_0} = C_{v_0}$, and
similar arguments can be used to prove the other two cases. After
inserting an edge $(u_0, v_0)$, by Lemma~\ref{lem:updatecore}, the
core number of every node in $V_c$ increases by at most 1. As a
result, after an edge $(u_0, v_0)$ insertion, for a node $w$ in
$V_c$, if $X_w < c + 1$, then $C_w$ will not increase. $C_w$ does
not change implying that $w$ is still in the $c$-core after
inserting an edge $(u_0, v_0)$. Clearly, it does not affect those
nodes in $V_c$ whose core numbers will increase to $c + 1$.
Therefore, we can safely remove the node $w$ from $V_c$. After
removing $w$, for any node $u \in V_c \backslash \{w\}$ that cannot
be reached from $u_0$ or $v_0$, we also can safely remove it from
$V_c$. The reason is because only the core number of the nodes that
are reachable from $u_0$ or $v_0$ may need to be updated. As a
consequence, for any node $u \in V_c$ such that every path either
from $u_0$ to $u$ or from $v_0$ to $u$ must go through $w$ can be
pruned. This completes the proof.
\end{myproof}
\eop

\begin{algorithm}[t]
\caption{void \textbf{XPruneColor}($G$, $u$, $c$)}
\label{alg:XPruneColor} {\small
\begin{algorithmic}[1]
\STATE visited($u$) $\leftarrow 1$;
\STATE $X_u \leftarrow 0$;
\FOR{each node $w \in N(u)$}
    \IF{$C_w \ge c$}
        \STATE $X_u \leftarrow X_u + 1$;
    \ENDIF
\ENDFOR
\IF {$X_u > c$}
    \IF {color($u$) = 0}
        \STATE $V_c \leftarrow V_c \cup \{u\}$;
        \STATE color($u$) = 1;
    \ENDIF
    \FOR {each node $w \in N(u)$}
        \IF {visited($w$) $= 0$ and $C_w = c$}
            \STATE \textbf{XPruneColor}($G$, $w$, $c$);
        \ENDIF
    \ENDFOR
\ENDIF
\end{algorithmic}
}
\end{algorithm}

Based on Theorem~\ref{thm:xpruninsert}, we can prune certain nodes
in the coloring procedure (the \textbf{Color} algorithm). We present
our new coloring algorithm with $X$-pruning in
Algorithm~\ref{alg:XPruneColor}. The new coloring algorithm is still
a DFS algorithm. The algorithm first calculates $X_u$ when it visits
a node $u$ (line~2-5 in Algorithm~\ref{alg:XPruneColor}). Based on
Theorem~\ref{thm:xpruninsert}, the DFS algorithm can early terminate
if it visits a node $u$ such that $X_u \le c$. The reason is that we
can safely remove such a node $u$ from $V_c$ by
Theorem~\ref{thm:xpruninsert}. Hence, the algorithm does not need to
recursively visits its neighbors. If $X_u > c$, the algorithm adds
node $u$ into $V_c$ and color it by 1 (line~7-9 in
Algorithm~\ref{alg:XPruneColor}). And then, the algorithm
recursively finds $u$'s neighbors in $V_c$ (line~10-12 in
Algorithm~\ref{alg:XPruneColor}). To implement this pruning
strategy, we can replace the \textbf{Color} algorithm with the
\textbf{XPruneColor} algorithm in Algorithm~\ref{alg:insertion}. 

\comment{
\begin{example}
  \label{exp:xpruninsert}Consider an example in
  Fig.~\ref{fig:kcoreupdate}. Assume that we insert an edge $(v_8,
  v_{10})$ into a graph. The algorithm will sequentially finds the
  nodes $v_8, v_{10}, v_9, v_2, v_{18}, v_{11}, v_{13}$. The nodes
  $v_1$ and $v_{12}$ are pruned, because $X_{v_1} = X_{v_{12}}=2 < 3$.
\end{example}
}

Second, we consider the edge deletion case. Suppose that we delete
an edge $(u_0, v_0)$ from graph $G$ and the core numbers of all the
nodes in $V_c$ are $c$. We consider three different cases: (1)
$C_{u_0} > C_{v_0}$, (2) $C_{u_0} < C_{v_0}$, and (3) $C_{u_0} =
C_{v_0}$. For $C_{u_0} > C_{v_0}$, we only need to find the nodes in
$G_{v_0}$, because the deletion of edge $(u_0, v_0)$ does not affect
the core number of the nodes in $G_{u_0}$. Recall that after
deleting an edge, the core number of the nodes in $V_c$ decreases by
at most 1. Therefore, after deleting an edge $(u_0, v_0)$, if
$X_{v_0} \ge c$, then $v_0$'s core number will not be changed. This
is because $X_{v_0} \ge c$ implies $v_0$ has at least $c$ neighbors
whose core numbers are larger than or equal to $c$. That is to say,
the core number of node $v_0$ is still $c$. Since $v_0$'s core
number does not change, we do not need to update the core number of
the nodes in $G_{v_0}$. As a result, under the case of $C_{u} >
C_{v}$ in Algorithm~\ref{alg:deletion} (line~4 in
Algorithm~\ref{alg:deletion}), we can first compute $X_{v}$. If $X_v
\ge c$, we do nothing. Symmetrically, for $C_{u_0} < C_{v_0}$, we
have a similar pruning rule as the case of $C_{u_0}
> C_{v_0}$. Also, for $C_{u_0} = C_{v_0}$, we first compute
$X_{u_0}$ and $X_{v_0}$. If $X_{u_0} < c$, then we need to update
the core number of the nodes in $G_{u_0}$. Also, if $X_{v_0} < c$,
we update the core number of the nodes in $G_{v_0}$. For the case
that $X_{u_0} \ge c$ and $X_{v_0} \ge c$, we do nothing, because no
node's core number needs to be updated. It is worth mentioning that
$X_{u_0}$ and $X_{v_0}$ are computed based on the core numbers of
the nodes that have not been updated. The detailed algorithm with
$X$-pruning for the edge deletion case is outlined in
Algorithm~\ref{alg:xprundeletion}. We can use the
\textbf{XPruneDeletion} algorithm to replace the \textbf{Deletion}
algorithm. The following example illustrates how this algorithm
works.

\begin{example}
  \label{exp:xprundeletion}Let us reconsider the example given in
  Fig.~\ref{fig:kcoreupdate}. Assume that we delete the dashed
  line (edge $(v_8, v_{10})$). In this case, the core number of $v_8$ and
  $v_{10}$ is 3. That is, $c = 3$. Then, we can calculate that $X_{v_8} = 2$ and $X_{v_{10}} =
  2$. Because $v_8$ has two neighbors ($v_5$ and $v_6$) whose core
  number is 4 and $v_{10}$ has two neighbors ($v_9$ and $v_{18}$)
  whose core numbers are 3. Since $X_{v_8} < c$ and $X_{v_{10}} < c$, we need to update the core number of the nodes in
  $G_{v_8}$ and $G_{v_{10}}$. After invoking
  Algorithm~\ref{alg:xprundeletion}, we can find that the core
  number of nodes $\{v_8, v_{10}, v_{9}, v_2, v_{18}\}$ decreases to 2.
\end{example}

\begin{algorithm}[t]
\caption{ \textbf{XPruneDeletion}($G$, $u$, $v$)}
\label{alg:xprundeletion} {\small
\begin{tabbing}
    {\bf\ Input}: \hspace{0.3cm}\= Graph $G=(V, E)$ and an edge $(u, v)$ \\
{\bf\ Output}: the updated core number of the nodes\>
\end{tabbing}
\begin{algorithmic}[1]
\STATE Initialize visited($w$) $\leftarrow 0$ for all node $w \in
V$; \STATE Initialize color($w$) $\leftarrow 0$ for all node $w \in
V$; \STATE $V_c \leftarrow \emptyset$; \STATE Compute $X_u$; \STATE
Compute $X_v$; \IF {$C_u$  $> C_v$}
    \STATE $c \leftarrow C_v$;
    \IF{$X_v < c$}
        \STATE \textbf{Color}($G$, $v$, $c$);
        \STATE \textbf{RecolorDelete}($G$, $c$);
        \STATE \textbf{UpdateDelete}($G$, $c$);
    \ENDIF
\ENDIF \IF {$C_u$  $< C_v$}
    \STATE $c \leftarrow C_u$;
    \IF{$X_u < c$}
        \STATE \textbf{Color}($G$, $u$, $c$);
        \STATE \textbf{RecolorDelete}($G$, $c$);
        \STATE \textbf{UpdateDelete}($G$, $c$);
    \ENDIF
\ENDIF \IF {$C_u$  $= C_v$}
    \STATE $c \leftarrow C_u$;
    \IF{$X_u < c$ and $X_v < c$}
        \STATE \textbf{Color}($G$, $u$, $c$);
        \IF{color($v$) $= 0$}
            \STATE Initialize visited($w$) $\leftarrow 0$ for all node $w \in V$;
            \STATE \textbf{Color}($G$, $v$, $c$);
            \STATE \textbf{RecolorDelete}($G$, $c$);
            \STATE \textbf{UpdateDelete}($G$, $c$);
        \ELSE
            \STATE \textbf{RecolorDelete}($G$, $c$);
            \STATE \textbf{UpdateDelete}($G$, $c$);
        \ENDIF
    \ENDIF
    \IF{$X_u < c$ and $X_v \ge c$}
        \STATE \textbf{Color}($G$, $u$, $c$);
        \STATE \textbf{RecolorDelete}($G$, $c$);
        \STATE \textbf{UpdateDelete}($G$, $c$);
    \ENDIF
    \IF{$X_u \ge c$ and $X_v < c$}
        \STATE \textbf{Color}($G$, $v$, $c$);
        \STATE \textbf{RecolorDelete}($G$, $c$);
        \STATE \textbf{UpdateDelete}($G$, $c$);
    \ENDIF
\ENDIF
\end{algorithmic}
}
\end{algorithm}

\stitle{$Y$-pruning: }For a node $w$, $Y_w$ is a lower bound of
$C_w$ by Lemma~\ref{lem:xylemma}. Here we develop pruning technique
using such lower bound, and we refer to this pruning technique as
$Y$-pruning.

To illustrate our idea, let us reconsider the toy induced core
subgraph shown in Fig.~\ref{fig:toygraph} which includes three
parts, $S_1$, $w$, and $S_2$. Suppose that we insert or delete an
edge $(u_0, v_0)$. Below, we focus on the case of $C_{u_0} =
C_{v_0}=c$, and similar descriptions can be used for other two
cases. Further, we assume that both $u_0$ and $v_0$ are in $S_1$,
and $Y_w = c$. First, we consider the insertion case, i.e., an edge
$(u_0, v_0)$ insertion. In this case, we claim that the core number
of the nodes in $S_2$ are unchanged. The reason is as follow. Let
$u$ in $S_2$ be a neighbor node of $w$. Then, for any neighbor $u$,
we have $Y_u < c$ (if not, $u$ and $w$ will be in a $(c+1)$-core).
This implies that for each neighbor of $w$ in $S_2$, the core number
cannot increase to $c+1$ after inserting $(u_0, v_0)$. As a result,
the core numbers of all the nodes in $S_2$ will not change after
inserting $(u_0, v_0)$. Second, for the deletion case, if we delete
an edge $(u_0, v_0)$, $C_w$ still equals to $c$ because $w$ has $c$
neighbors whose core numbers are larger than $c$ ($Y_w=c$). Clearly,
the core numbers of the nodes in $S_2$ are also unchanged. Put it
all together, under both edge insertion and edge deletion cases, the
core numbers of all the nodes in $S_2$ will not change, and thereby
we can safely prune the nodes in $S_2$. Formally, for $Y$-pruning,
we have the following theorem.

\begin{theorem}
  \label{thm:ypruneinsert}Given a graph $G$ and an edge $(u_0, v_0)$. After inserting/deleting an edge $(u_0, v_0)$ in $G$,
  for a node $w \in V_c$, if $Y_w = c$, then we have the following
  pruning rules.
\begin{itemize}
  \item If $C_{u_0} > C_{v_0}$ (i.e., $V_c = V_{{v_0}}$), then for
  any node $u \in V_c$ and $u \ne w$ that every path from $v_0$ to $u$ must go
  through $w$ can be pruned.

  \item If $C_{u_0} < C_{v_0}$ (i.e., $V_c = V_{{u_0}}$), then for
  any node $u \in V_c$ and $u \ne w$ that every path from $u_0$ to $u$ must go
  through $w$ can be pruned.

  \item If $C_{u_0} = C_{v_0}$ (i.e., $V_c = V_{{u_0 \cup v_0}}$),
  then for any node $u \in V_c$ and $u \ne w$ that every path either from $u_0$ to $u$ or from $v_0$
  to $u$ must go through $w$ can be pruned.
\end{itemize}
\end{theorem}

\begin{myproof}
 We prove this theorem under the case $C_{u_0} = C_{v_0}$, and for
 other cases, we have similar proofs. 
 Below, we discuss the proofs for the edge insertion and edge deletion cases, respectively.

First, we prove the edge insertion case.
 Let $V_{>c}$ be a set of nodes whose core
 numbers are larger than $c$. Assume that we remove $w$ from $V_c$. Then,
 after removing $w$, we denote a set of nodes in $V_c$ that cannot be reachable either from
$u_0$ or from $v_0$ as $V_1$. Then, after inserting an edge $(u_0,
v_0)$, we consider two cases: (1) $w$'s core number will not change,
and (2) $w$'s core number increases by 1. The first case suggests
that $w$ is still in the $c$-core, and we can safely remove $w$ from
$V_c$. Therefore, for the nodes in $V_1$, we can also remove them
from $V_c$, because only the core number of those nodes that are
reachable from $u_0$ or $v_0$ may need to be updated. Second, we
consider the case that $w$'s core number increases by 1 after
inserting an edge $(u_0, v_0)$. We denote a subset of nodes in $V_c$
whose core numbers increase by 1 as $\tilde V_c$ after inserting an
edge $(u_0, v_0)$. Further, we denote a subset of nodes in $V_1$
whose core numbers need to increase by 1 as $V_2$. In other words,
$V_2 = V_1 \bigcap \tilde V_c$. Clearly, the theorem holds if $V_2 =
\emptyset$. Now we prove this by contradiction. Specifically, we
assume that $V_2 \ne \emptyset$. By definition, after inserting an
edge $(u_0, v_0)$, the induced subgraph by the nodes in $\tilde V_c
\bigcup V_{>c}$ forms a $(k+1)$-core. We denote such subgraph as $G
^\prime = (V ^\prime, E^\prime)$, where $V ^\prime = \tilde V_c
\bigcup V_{>c}$. Clearly, all the nodes in $G ^\prime$ has  at least
a degree $c + 1$. Now consider a subgraph $G ^ \star$ induced by the
nodes in $V_2 \bigcup \{w\} \bigcup V_{>c}$. We claim that all the
nodes in $G ^ \star$ has at least a degree $c + 1$. First, for the
nodes in $V_{>c}$, their degree is obviously greater than $c + 1$
w.r.t.\ $G ^ \star$. Second, we consider the nodes in $V_2$. By
definition, in graph $G ^\prime$, there is no edge between the nodes
in $V_2$ and the nodes in $\tilde V_c \backslash \{V_2 \bigcup
\{w\}\}$. Since the nodes in $V_2$ have at least a degree $c + 1$
w.r.t.\ graph $G ^\prime$, they also have at least a degree $c + 1$
w.r.t.\ graph $G ^\star$. Third, we consider the node $w$. On the
one hand, we claim that $w$ has at least one neighbor in $V_2$.
Suppose $w$ has no neighbor in $V_2$, then the nodes in $V_2$ whose
core numbers cannot increase to $c+1$ after inserting an edge $(u_0,
v_0)$ by the $k$-core update theorem, which contradict to our
assumption. Hence, $w$ has at least one neighbor in $V_2$. On the
other hand, since $Y_w = c$, $w$ has $c$ neighbors whose core
numbers are larger than $c$. As a result, $w$ has at least a degree
$c + 1$ w.r.t.\ graph $G ^\star$. Put it all together, all the nodes
in $G ^\star$ have at least a degree $c + 1$. Note that by our
definition the induced subgraph $G ^\star$ does not contain node
$u_0$ and $v_0$. Consequently, before inserting the edge $(u_0,
v_0)$, the core number of the nodes in $G ^\star$ at least $c + 1$.
That is to say, the nodes in $V_2$ has core number $c + 1$ before
inserting the edge $(u_0, v_0)$, which is a contradiction. This
completes the proof for the edge insertion case.

For the edge deletion case, after deleing an edge $(u_0, v_0)$, the
core number of all the nodes in $V_c$ decreases by at most 1
according to Lemma~\ref{lem:updatecore}. Hence, if a node $w \in
V_c$ has $Y_w = c$, then $w$'s core number will not decrease.
Similarly, let $V_{>c}$ be a set of nodes whose core
 numbers are larger than $c$. And assume that we remove $w$ from $V_c$. Then, after removing $w$, we denote a
set of nodes in $V_c$ that cannot be reachable either from $u_0$ or
from $v_0$ as $V_1$. Now consider a subgraph $G ^{\star}$ induced by
the nodes $V_1 \bigcup \{w\} \bigcup V_{>c}$. We claim that all the
nodes in such subgraph have at least a degree $c$. First, for the
nodes in $V_{>c}$, their degree is clearly larger than $c$ w.r.t.\
$G ^{\star}$ because their core numbers are larger than $c$. Second,
$w$'s degree is at least $c$ w.r.t.\ $G ^{\star}$, because $w$ has
$c$ neighbors whose core numbers are larger than $c$. Third, for the
nodes in $V_1$, their degree is also at least $c$ w.r.t.\ $G
^{\star}$. The rationale is as follows. By definition, no edge in
$G$ goes through the nodes in $V_c \backslash \{V_1 \cup \{w\}\}$
and the nodes in $V_1$. Since the core number of the nodes in $V_1$
is $c$, the nodes in $V_1$ has at least $c$ neighbors w.r.t.\ $G
^{\star}$. Consequently, the core number of the nodes in $G
^{\star}$ is still $c$ after removing the edge $(u_0, v_0)$. This
implies that the nodes in $V_1$ can be pruned, which completes the
proof for the edge deletion case.
\end{myproof}
\eop

\begin{algorithm}[t]
\caption{void \textbf{YPruneColor}($G$, $u$, $c$)}
\label{alg:YPruneColor} \small
\begin{algorithmic}[1]
\STATE visited($u$) $\leftarrow 1$; \IF {color($u$) = 0}
        \STATE $V_c \leftarrow V_c \cup \{u\}$;
        \STATE color($u$) = 1;
\ENDIF
\STATE $Y_u \leftarrow 0$;
\FOR{each node $w \in N(u)$}
    \IF{$C_w > c$}
        \STATE $Y_u \leftarrow Y_u + 1$;
    \ENDIF
\ENDFOR
\IF {$Y_u < c$}
    \FOR {each node $w \in N(u)$}
        \IF {visited($w$) $= 0$ and $C_w = c$}
            \STATE \textbf{YPruneColor}($G$, $w$, $c$);
        \ENDIF
    \ENDFOR
\ENDIF
\end{algorithmic}
\end{algorithm}

Based on Theorem~\ref{thm:ypruneinsert}, we can implement the
$Y$-pruning strategy in the coloring procedure. We present our new
coloring algorithm with $Y$-pruning in
Algorithm~\ref{alg:YPruneColor}, which is also a DFS algorithm. In
particular, Algorithm~\ref{alg:YPruneColor} first colors a node $u$
by 1 and adds it into $V_c$ when it visits $u$ (line~2-4 in
Algorithm~\ref{alg:YPruneColor}). Then, the algorithm calculates
$Y_u$ (line~5-8 in Algorithm~\ref{alg:YPruneColor}). If $Y_u = c$,
then the algorithm can early terminate. The reason is because the
nodes that cannot be reachable from $u_0$ or $v_0$ after removing
$u$ can be pruned by Theorem~\ref{thm:ypruneinsert}. If $Y_u < c$,
the algorithm recursively finds $u$'s neighbors in $V_c$ (line~9-12
in Algorithm~\ref{alg:YPruneColor}). Below, we discuss how to
integrate the \textbf{YPruneColor} algorithm into the
\textbf{Insertion} and \textbf{Deletion} algorithm.

First, to integrate the \textbf{YPruneColor} algorithm into the
\textbf{Insertion} algorithm, we need to replace the \textbf{Color}
algorithm with the \textbf{YPruneColor} algorithm as well as handle
the following special case. That is, if $C_{u_0} = C_{v_0} = c$,
$Y_{u_0} = c$ and $Y_{v_0} < c$, we need to invoke
\textbf{YPruneColor}($G$, $v_0$, $c$). If $C_{u_0} = C_{v_0} = c$,
$Y_{u_0} < c$ and $Y_{v_0} = c$, we need to invoke
\textbf{YPruneColor}($G$, $u_0$, $c$). The reason is because we need
to allow the DFS algorithm to go through the edge $(u_0, v_0)$ in
order to add both $u_0$ and $v_0$ into $V_c$. If $C_{u_0} = C_{v_0}
= c$ and $Y_{u_0} = Y_{v_0} = c$, then we have to invoke both
\textbf{YPruneColor}($G$, $u_0$, $c$) and \textbf{YPruneColor}($G$,
$v_0$, $c$) so as to add both $u_0$ and $v_0$ into $V_c$. Second, to
integrate the  \textbf{YPruneColor} algorithm into the
\textbf{Deletion} algorithm, we only need to replace the
\textbf{Color} algorithm with the \textbf{YPruneColor} algorithm.
The following example illustrates how the \textbf{YPruneColor}
algorithm works.

\begin{example}
  \label{exp:ypruninsert} Consider an example in
  Fig.~\ref{fig:kcoreupdate}. For the edge insertion case, we assume that the edge $(v_8,
  v_{10})$ is the inserted edge. Since
  $v_8=v_{10}=c=2$ and $Y_{v_8}=2$, we invoke \textbf{YPruneColor}($G$, $v_{10}$,
  $2$). The algorithm first colors $v_{10}$ by 1 and adds it into
  $V_c$. Then, the algorithm colors node $v_8$ by 1 and adds it into
  $V_c$. Since $Y_{v_8}=2$, the recursion terminates at $v_8$ and
  returns to $v_{10}$. Similarly, when the algorithm visits node
  $v_2$, the recursion also terminates as $Y_{v_2}=2$. As a result,
  the node $v_1$ is pruned. Finally, we can obtain $V_c=\{ v_{10}, v_8, v_9, v_3, v_{18}, v_{11}, v_{12}, v_{13} \}$ after the algorithm
  ends.

  For the edge deletion case, we also assume that we delete an edge $(v_8,
  v_{10})$ from $G$. Under this case, we have $v_8=v_{10}=c=3$.
  Since no node in $V_c$ has $Y_u = c$, the
  $Y$-pruning cannot prune any node. Suppose that the edge $(v_8,
  v_{10})$ is deleted. Then, we have $v_9 = v_{10} = c =
  2$. Under this case, assume that we further delete an edge $(v_9,
  v_{10})$. Then, we can find that the set $V_c$ contains nodes $\{v_9, v_2, v_1, v_{10},
v_{18}, v_{11}, v_{12}, v_{13}\}$. Since $Y_{v_2}=c=2$, the node
$v_1$ can be pruned by the \textbf{YPruneColor} algorithm.
\end{example}

\stitle{Combination of $X$-pruning and $Y$-pruning: }Here we discuss
how to combine both $X$-pruning and $Y$-pruning for edge insertion
case and edge deletion case, respectively. For edge insertion case,
we can integrate both $X$-pruning and $Y$-pruning into the coloring
procedure. Specifically, in the coloring procedure, when the DFS
algorithm visits a node $u$, we calculate both $X_u$ and $Y_u$.
Then, we use the $X$-pruning rule to determine the color of node
$u$, and make use of both $X$-pruning and $Y$-pruning rules to
determine whether the algorithm needs to recursively visits $u$'s
neighbors or not. For edge insertion, the detailed coloring
algorithm with both $X$-pruning and $Y$-pruning, called
\textbf{XYPruneColor}, is outlined in
Algorithm~\ref{alg:XPruneColor}.

For the edge deletion case, we can easily integrate both $X$-pruning
and $Y$-pruning via the following two steps. First, we replace the
\textbf{Color} algorithm in \textbf{Deletion} with the
\textbf{YPruneColor} algorithm. Second, we integrate the $X$-pruning
rule into the \textbf{Deletion} algorithm. First, we replace the
\textbf{Color} algorithm in \textbf{XPrunDeletion} with the
\textbf{YPruneColor} algorithm. Second, we use this
\textbf{XPrunDeletion} algorithm to replace the \textbf{Deletion}
algorithm.

\begin{algorithm}[t]
\caption{void \textbf{XYPruneColor}($G$, $u$, $u_0$, $c$)}
\label{alg:xYPruneColor} {\small
\begin{algorithmic}[1]
\STATE visited($u$) $\leftarrow 1$; \STATE $X_u \leftarrow 0$;
\STATE $Y_u \leftarrow 0$; \FOR{each node $w \in N(u)$}
    \IF{$C_w \ge c$}
        \STATE $X_u \leftarrow X_u + 1$;
    \ENDIF
    \IF{$u \ne u_0$ and $C_w > c$}
        \STATE $Y_u \leftarrow Y_u + 1$;
    \ENDIF
\ENDFOR \IF{$X_u > c$}
    \IF {$Y_u < c$ or $c = 0$}
        \FOR {each node $w \in N(u)$}
            \IF {visited($w$) $= 0$ and $C_w = c$}
                \STATE \textbf{XYPruneColor}($G$, $w$, $u_0$, $c$);
            \ENDIF
        \ENDFOR
    \ENDIF
    \IF {color($u$) = 0}
            \STATE $V_c \leftarrow V_c \cup \{u\}$;
            \STATE color($u$) = 1;
    \ENDIF
\ENDIF
\end{algorithmic}
}
\end{algorithm}

\vspace*{-0.2cm}
\section{Experiments}
\label{sec:experiments} In this section, we conduct comprehensive
experiments to evaluate our approach. In the following, we first
describe our experimental setup and then report our results.

\subsection{Experimental setup}
\label{subsec:expsetup}

\stitle{Different algorithms: }We compare 5 algorithms. The first
algorithm is the baseline algorithm, which invokes the $O(n + m)$
algorithm to update the core number of nodes given the graph is
updated \cite{10kcoredynamic}. We denote this algorithm as algorithm
B. The second algorithm is our basic algorithm without pruning
strategies, which is denoted as algorithm N. The third algorithm is
our basic algorithm with $X$-pruning, which is denoted as algorithm
X. The fourth algorithm is our basic algorithm with $Y$-pruning,
which is denoted as algorithm Y. The last algorithm is our basic
algorithm with both $X$-pruning and $Y$-pruning, which is denoted as
algorithm XY.

\stitle{Datasets: }We collect 15 real-world datasets to conduct our
experiments. Our datasets are described as follows. (1)
Co-authorship networks: we download four physics co-authorship
networks from Stanford network data collections \cite{standforddata}
which are HepTh, HepPh, Astroph, and CondMat datasets. In addition,
we also extract a co-authorship network from a subset of the DBLP
dataset (\url{www.informatik.uni-trier.de/~ley/db}) with 78,649
authors. (2) Online social networks: we collect the Douban
(\url{www.douban.com}) dataset from ASU social computing data
repository \cite{asudatasets}, and collect the Epinions
(\url{www.epinions.com}), two Slashdot datasets
(\url{www.slashdot.org}), and the Wikivote dataset from Stanford
network data collections \cite{standforddata}. (3) Communication
networks: we employ two Email communication networks, namely
EmailEnron and EmailEuAll, from Stanford network data collections
\cite{standforddata}. (4) P2P networks: we download a P2P network
(Gnutella) dataset from Stanford network data collections
\cite{standforddata}, which are originally collected from Gnutella
\cite{standforddata}. (5) Location-based social networks (LBSNs): We
download two notable LBSNs datasets from Stanford network data
collections \cite{standforddata}. For all the datasets, if the graph
is a directed graph, we ignore the direction of the edges in the
graph. The detailed statistical information of our datasets are
described in Table~\ref{tbl:data}.

\begin{table}[t]
\begin{center}\vspace*{-0.3cm}\small
\caption[]{Summary of the datasets} \label{tbl:data}
\begin{tabular}{l|c|c|c|c}
\hline
Name & \#nodes & \#edges & Ref. & Description \\
\hline \hline
HepTh & 9,877  & 51,946 & \cite{standforddata} &   \\
HepPh & 12,008 & 236,978 & \cite{standforddata} & Co-authorship \\
Astroph & 18,772 & 396,100 & \cite{standforddata} & networks\\
CondMat & 23,133  & 186,878 & \cite{standforddata} &  \\
DBLP & 78,649 & 382,294 & website &  \\
\hline
Douban & 154,908 & 654,324 & \cite{asudatasets} & \\
Epinions & 75,872 & 396,026 & \cite{standforddata} & Online   \\
Slashdot1 & 77,360 & 826,544 & \cite{standforddata} & social  \\
Slashdot2 & 82,168 & 867,372 & \cite{standforddata} & networks\\
Wikivote & 5,311 & 142,066 & \cite{standforddata} & \\
\hline
EmailEnron & 36,692  & 367,662 & \cite{standforddata} &  Communication\\
EmailEuAll & 265,182 & 224,372 & \cite{standforddata} &  networks\\
\hline
Gnutella & 62,586 & 153,900 &  \cite{standforddata} & P2P networks\\
\hline
Brightkite & 58,228 & 428,156 & \cite{standforddata} & Location based \\
Gowalla & 196,591 & 1,900,654 & \cite{standforddata} & social networks\\
\hline
\end{tabular}\vspace*{-0.5cm}
\end{center}
\end{table}

\stitle{Experimental environment: }We conduct our experiments on a
Windows Server 2007 with 4xDual-Core Intel Xeon 2.66 GHz CPU, and
128G memory. All the algorithms are implemented by Visual C++ 6.0.

\subsection{Results for single edge updates}
\label{subsec:expresults}For all the experiments, we randomly delete
and insert 500 edges in the original datasets. After
inserting/deleting an edge, we invoke 5 different algorithms to
update the core number of the nodes, respectively. For all the
algorithms, we record the average time to update the core number of
nodes over 500 edge insertions and 500 edge deletions. Specifically,
we record three quantities, namely average insertion time, average
deletion time, and average update time. We calculate the average
insertion (deletion) time by the average core number update time of
different algorithms over 500 edge insertions (deletions). The
average update time is the mean of average insertion time and
average deletion time. To evaluate the efficiency of our algorithms
(algorithm N, algorithm X, algorithm Y, algorithm XY), we compare
them with the baseline algorithm (algorithm B) according to the
average insertion/deletion/update time. Our results are depicted in
Table~\ref{tbl:result}.

\begin{table*}[t]
\begin{center}{\small
\caption[]{Average update time of different algorithms (In last column, SR denotes the speedup ratio of XY). All time is millisecond.} 
\label{tbl:result}
\begin{tabular}{l|c|c|c|c|c|c|c|c|c|c|c|c|c|c|c|c}
\hline \multirow{2}{*}{Time (ms)} & \multicolumn{5}{|c|}{Average
deletion time} & \multicolumn{5}{|c|}{Average insertion time} &
\multicolumn{6}{|c}{Average update time}  \\ \cline{2-17} &
B & N & X & Y & XY & B & N & X & Y & XY & B & N & X & Y & XY & SR\\
\hline \hline
HepTh       &  2.38  & 1.06  & 0.54  & 1.00  & \textbf{0.48}  &  2.80 &  1.32 & 1.20  & 1.28  & \textbf{1.14}  &  2.59 & 1.19  &  0.87 & 1.14  & \textbf{0.81} & 3.2 \\
HepPh       &  4.12  & 2.58  & 1.30  & 1.58  & \textbf{1.20}  &  5.30 &  1.46 & 1.32  & 1.40  &  \textbf{1.20} &  4.71 & 2.02  &  1.31 & 1.49  & \textbf{1.20} & 3.9 \\
Astroph     &  9.14  & 1.30  & 0.36  & 1.12  & \textbf{0.32}  &  9.92 &  1.56 &  1.40 & 1.42 & \textbf{1.40}  &  9.53 & 1.43  &  0.88 & 1.27  & \textbf{0.86} & 11.1\\
CondMat     &  5.94  & 1.52  & 0.64  & 1.30  & \textbf{0.60}  &  6.24 &  1.50 &  1.40 &  1.36 & \textbf{1.32}  &  6.09 & 1.51  & 1.02  & 1.33  & \textbf{0.96} & 6.3\\
DBLP        &  12.08 & 1.68  & 1.26  & 1.48  & \textbf{1.22}  &  12.22 &  1.52 &  1.42 &  1.44 & \textbf{1.38}  &  12.15 & 1.60  & 1.34  & 1.46  & \textbf{1.30} & 9.3\\
\hline
Douban      &  21.38 & 4.58  & 2.14  & 3.28  & \textbf{1.32}  & 21.16 & 2.62  & 2.02  & 2.40  & \textbf{2.00}  &  21.27 & 3.60  & 2.08  & 2.84  & \textbf{1.66} & 12.8\\
Epinions    &  13.00 & 2.06  & 0.68  & 1.62  & \textbf{0.64}  & 13.94  & 2.04  & 1.56  & 1.80  & \textbf{1.50}  & 13.47  & 2.05  & 1.12  & 1.71  & \textbf{1.07} & 12.6\\
Slashdot1   &  22.53 & 4.12  & 1.43  & 2.06  & \textbf{1.38}  & 20.37  & 2.80  & 1.73  & 1.88  & \textbf{1.32}  & 20.45  & 3.46  & 1.58  & 1.87  & \textbf{1.35} & 15.1\\
Slashdot2   &  24.36 & 4.85  & 1.56  & 2.13  & \textbf{1.54}  & 22.32  & 2.93  & 1.82  & 2.05  & \textbf{1.64}  & 23.34  & 3.73  & 1.69  & 2.09  & \textbf{1.59} & 14.7\\
Wikivote    &  3.64 &  1.32  & 0.50  & 0.50  &  \textbf{0.48} &  4.06 & 1.78  & 1.70  & 1.76 &  \textbf{1.42}  &  3.85 & 1.55  &  1.10  & 1.13  & \textbf{0.95} & 4.1\\
\hline
EmailEnron  &  10.80 &  2.40  & 0.90  &  1.82 &  \textbf{0.86} &  10.60 &  2.92 & 2.70  & 2.82  & \textbf{2.68}  &  10.70 & 2.66  &  1.80 & 2.32  & \textbf{1.77} & 6.0\\
EmailEuAll  &  13.06 &  2.14  & 1.24  & 1.64  & \textbf{1.22}  &  12.52 & 1.74  &  1.52 & 1.70  & \textbf{1.24}  & 12.79  & 1.94  &  1.38 & 1.67  & \textbf{1.23}& 10.4\\
\hline
Gnutella    &  10.32 & 2.64   & 1.58  & 1.66  & \textbf{1.38}  & 12.08  & 2.18  & 2.06  & 2.12  & \textbf{1.82}  & 11.20  & 2.41  & 1.82  & 1.89  & \textbf{1.60} & 7.0\\
\hline
Brightkite  &  13.60 &  1.56  & 0.64  & 1.32  & \textbf{0.54}  & 13.64  & 1.64  & 1.32  & 1.34  &  \textbf{1.32} & 13.62  & 1.60  & 0.98  & 1.33  & \textbf{0.93} & 14.6\\
Gowalla     &  108.20 &  2.10  &  1.12 &  1.82 &  \textbf{0.91} &  107.52 & 1.74  &  1.52 &  1.64 & \textbf{1.21}  &  107.86 &  1.92 & 1.32  &  1.73 & \textbf{1.06} & 101.8\\
\hline
\end{tabular}}\vspace*{-0.5cm}
\end{center}
\end{table*}

From Table~\ref{tbl:result}, we can clearly see that all of our
algorithms (algorithm N, algorithm X, algorithm Y, algorithm XY)
perform much better than the baseline algorithm (algorithm B) over
all the datasets used. The best algorithm is the algorithm XY, which
is our basic algorithm with both $X$-pruning and $Y$-pruning,
followed by algorithm X, algorithm Y, algorithm N, and algorithm B.
Over all the datasets used, the maximal speedup of our algorithms is
achieved in Gowalla dataset (the last row in
Table~\ref{tbl:result}). Specifically, in Gowalla dataset, algorithm
XY, algorithm X, algorithm Y and algorithm N reduce the average
update time of algorithm B by 101.8, 81.7, 62.3, and 56.2 times,
respectively. The minimal speedup of our algorithms is achieved in
HepTh dataset (the first row in Table~\ref{tbl:result}). In
particular, in HepTh dataset, algorithm XY, algorithm X, algorithm Y
and algorithm N reduce the average update time of algorithm B by
3.2, 3.0, 2.3, and 2.2 times respectively. In general, we find that
the speedup of our algorithms increases as the graph size increases.
The reason is because the time complexity of the baseline algorithm
is linear w.r.t.\ the graph size for handling each edge
insertion/deletion. Instead, the time complexity of our algorithms
is independent of the graph size, and it is only depends on the size
of the induced core subgraph. Additionally, over all the datasets,
we can observe that our basic algorithm with pruning techniques is
significantly more efficient than the basic algorithm without
pruning techniques. Below, we discuss the effect of the $X$-pruning
and $Y$-pruning techniques.

\stitle{The effect of pruning: }Here we investigate the effective of
our pruning techniques. From Table~\ref{tbl:result}, over all the
datasets, we can see that the $X$-pruning strategy (algorithm X) is
more effective than the $Y$-pruning strategy (algorithm Y) according
to average deletion/insertion/update time. For example, in HepTh
dataset (row 1 in Table~\ref{tbl:result}), algorithm X reduces the
average deletion time, the average insertion time, and the average
update time, over algorithm N by 96.3\%, 10\%, and 36.8\%,
respectively. However, in HepTh dataset, algorithm Y reduces the
average deletion time, the average insertion time, and the average
update time, over algorithm N by 6\%, 3.1\%, and 4.3\%,
respectively. This result indicates that the condition of the
$Y$-pruning is stronger than the condition of the $X$-pruning in
many real graphs. Recall that by Theorem~\ref{thm:ypruneinsert}, if
there is at least one node $u$ with core number $C_u$ and $Y_u =
C_u$ in the induced core subgraph, then the $Y$-pruning strategy may
prune some nodes. The condition of $Y$-pruning strategy ($Y_u =
C_u$) is strong, because if a node has $C_u$ neighbors whose core
number is larger than $C_u$, then this node may have another
additional neighbor whose core number is larger than $C_u$, thus
resulting in that the node $u$ is in a $(C_u + 1)$-core. Instead,
indicating by our experimental result, the condition of the
$X$-pruning strategy ($X_u \le C_u + 1$) may be easily satisfied in
real graphs. This result also implies that the lower bound of the
core number in Lemma~\ref{lem:xylemma} ($Y_v$) is typically very
loose for many nodes in real graphs. In addition, we can observe
that the algorithm with both $X$-pruning and $Y$-pruning strategies
is more efficient than the algorithm with only one pruning strategy
over all the datasets. Generally, we find that the $X$-pruning
strategy under the edge deletion case is more effective than itself
under the edge insertion case. Similarly, the $Y$-pruning strategy
under the edge deletion case is more effective than itself under the
edge insertion case. Taking the Gnutella dataset as an example (row
13 in Table~\ref{tbl:result}), for the edge deletion case, algorithm
X reduces the average deletion time over algorithm N by 143.75\%,
while for the edge insertion case, algorithm X cuts the average
insertion time over algorithm N only by 5.8\%. For the edge deletion
case, algorithm Y reduces the average deletion time over algorithm N
by 59\%, while for the edge insertion case, algorithm Y reduces the
average insertion time over algorithm by 2.8\%.

\comment{ \stitle{Correctness of our algorithms: }To confirm the
correctness of our algorithms, we record the core number of every
node computed by different algorithms after deleting and inserting
500 edges. Then, for each node, we compare its core number
calculated by our algorithms (algorithm N, algorithm X, algorithm Y,
and algorithm XY) with its core number computed by the baseline
algorithm (B). Fig.~\ref{fig:correct} depicts the core number of all
the nodes in Wikivote datasets by different algorithms. Similar
results can be observed from other datasets. From
Fig.~\ref{fig:correct}, we can observe that all of our algorithms
generate the same figure as the baseline algorithm. These results
indicate that the core number calculated by all of our algorithms
consists with the core number calculated by the baseline algorithm.
This confirms the correctness of our algorithms.

\begin{figure*}[t]
\begin{center}
\includegraphics[width=\linewidth]{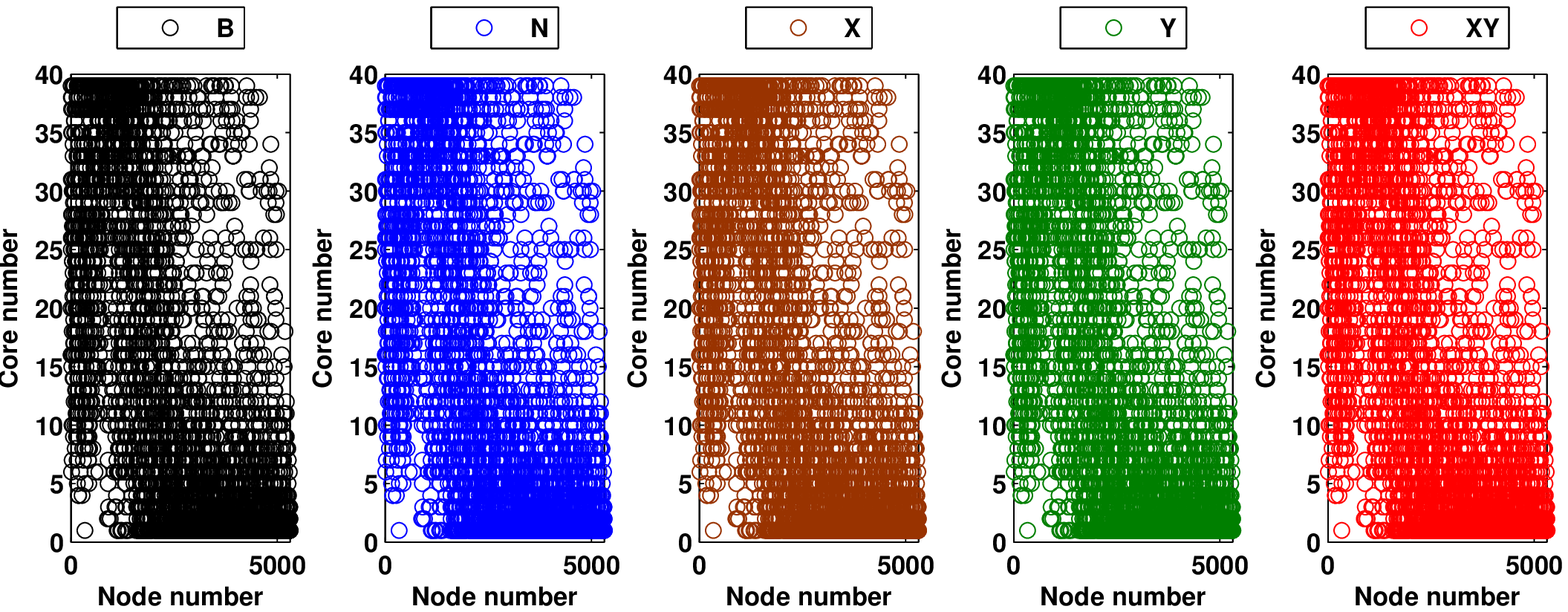}
\vspace*{-2em} \caption{Correctness of our algorithms.}
\label{fig:correct} \vspace*{-0.8cm}
\end{center}
\end{figure*}
}

\subsection{Results for a batch of edge updates}
\label{subsec:discuss}In previous experiments, we have shown the
performance of our algorithms for core maintenance in a graph given
the graph is updated by an edge insertion or deletion. These
algorithms are extremely useful to \emph{continuously} monitor the
dynamics of the core number of the nodes in time-evolving graph.
Besides the graph with a single edge update, here we show the
performance of our algorithms in a dynamic graph given a batch of
edges updates. Assume that the graph has $r$ edge updates at a time
interval $\Delta t$. To maintain the core number of the nodes, we
need to sequentially invoke our algorithm (algorithm XY) $r$ times.
For the baseline algorithm (algorithm B), however, we can invoke it
one time to recompute the core number of all nodes. Since our XY
algorithm is the best algorithm for single edge updates, we only
compare our XY algorithm with algorithm B.

Now, let us focus on the last column in Table~\ref{tbl:result} which
shows the speedup ratio (SR) of algorithm XY over algorithm B for a
single edge update. In general, if $r$ is less than the speedup
ratio, then our algorithm is more efficient than the baseline
algorithm for processing a batch of edge updates at a time interval
$\Delta t$. For example, in Gowalla dataset, the speedup ratio of
our algorithm is 101.8. As a result, if the graph has less than 101
edge updates, i.e., $r \le 101$, then our algorithm is more
efficient than the baseline algorithm. However, if $r$ is larger
than the speedup ratio of our algorithm, the baseline algorithm is
more preferable than our algorithm. As shown in
Table~\ref{tbl:result}, the speedup ratio of our algorithm increases
as the graph size increases. This result implies that, for a batch
of edge updates, our algorithm is very efficient in large graphs
with small $r$. In other words, if the graph is very large and
evolves slowly, then our algorithm is more preferable. However, if
the graph is very small and frequently varying, then the baseline
algorithm is more efficient than our algorithm. Below, we show the
speedup ratio of our algorithm in large synthetic graphs.

\begin{figure}[t]
\begin{center}
\includegraphics[scale=0.40]{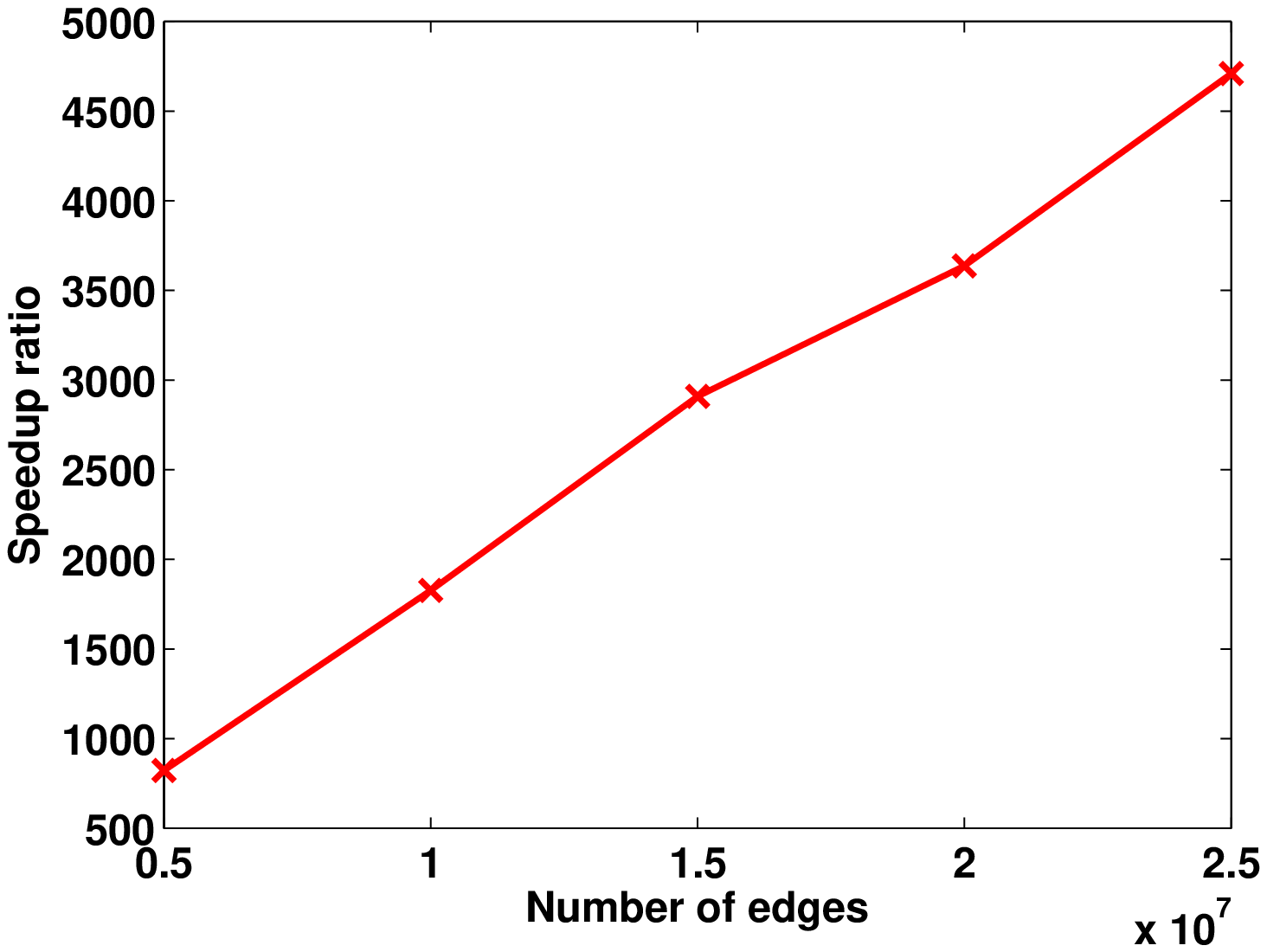}
\vspace*{-1em} \caption{Speedup ratio vs. graph size.}
\label{fig:spratio} \vspace*{-0.5cm}
\end{center}
\end{figure}

To evaluate the speedup ratio of our algorithm in large graphs, we
generate five large synthetic graphs based on a power-law random
graph model \cite{99sfnet}. Specifically, we produce five synthetic
graphs $G_1, \cdots, G_5$ with $G_i$ has $i$ million nodes and $5
\times i$ million edges for $i = 1, \cdots, 5$. Then, we adopt the
same method used in our previous experiments to compute the speedup
ratio of our algorithm. Fig.~\ref{fig:spratio} shows that the result
of speedup ratio of our algorithm with different graph size. From
Fig.~\ref{fig:spratio}, we can see that the speedup ratio is greater
than 4700 when the graph size is 5 million nodes and 25 million
edges. That is to say, in such a graph, if $r$ is smaller than 4700,
then our algorithm is more efficient than the baseline algorithm.
Generally, for a fixed graph size (from 1 million to 5 million
nodes), if $r$ is below the red curve in Fig.~\ref{fig:spratio},
then our algorithm is more preferable than the baseline algorithm,
otherwise the baseline algorithm is more efficient.

\vspace*{-0.2cm}
\section{Related work}
\label{sec:rlwork}The $k$-core decomposition in networks has been
extensively studied in the literature. In \cite{83kcoredef}, Seidman
introduces the concept of $k$-core for measuring the group cohesion
in a network. The cohesion of the $k$-core increases as $k$
increases. Recently, the $k$-core decomposition in graph has been
successfully used in many application domains, such as visualization
of large complex networks \cite{99kcoreviz, 04kcoredrawing,
05nipskcoreviz, 07corrkcoreviz, 12icdetrikcore}, uncovering the
topological structure of the Internet \cite{07pnaskcore,
08kcoreinternet, 11imkcoreconncetive}, analysis of the structure and
function of the biological networks \cite{03proteinkcore,
03proteinfunkcore, 05proteinkcore}, studying percolation in random
graph \cite{06prlkcore, 06kcorepercolation}, as well as identifying
the influential spreader in complex network \cite{10naturephykcore}.
Below, we list some notable work on these applications.

In \cite{99kcoreviz}, Batagelj et al.\ propose to use $k$-core
decomposition to visualize the large graph. Specifically, they first
partition a large graph into smaller parts using the $k$-core
decomposition and then visualize each smaller part by standard graph
visualization tools. In \cite{04kcoredrawing}, based on the $k$-core
decomposition, Baur et al.\ present a method for drawing autonomous
systems graph using 2.5D graph drawing. Their algorithm makes use of
a spectral layout technique to place the nodes in the highest order
core. Then, the algorithm uses an improved directed-forces method to
place the nodes in each $k$ core according to the decreasing order.
Alvarez-Hamelin et al.\ \cite{05nipskcoreviz, 07corrkcoreviz}
propose a visualization algorithm to uncover the hierarchical
structure of the network using $k$-core decomposition. Their
algorithm is based on the hierarchical property of $k$-core
decomposition. More recently, Zhang and Parthasarathy
\cite{12icdetrikcore} introduce a different notion, namely triangle
$k$-core, to  extract the clique-like structure and visualize the
graph. Unlike the traditional $k$-core, the triangle $k$-core is the
maximal subgraph that each edge of the subgraph is contained within
at least $k$ triangles. They also propose a maintenance technique
for triangle $k$-core. Since the triangle $k$-core is totally
different from $k$-core, their maintenance technique cannot be
applied in our problem. The $k$-core decomposition is also
successfully used for analyzing and modeling the structure of the
Internet \cite{07pnaskcore, 08kcoreinternet, 11imkcoreconncetive}.
For example, in \cite{07pnaskcore}, Carmi et al.\ study the problem
of mapping the Internet using the method of $k$-core decomposition.
In \cite{08kcoreinternet}, Alvarez-Hamelin et al.\ investigate the
hierarchies and self-similarity of the
Internet using $k$-core decomposition. 
Besides the Internet, the $k$-core decomposition has also been
applied to analyze the structure and function of the biological
networks. In \cite{03proteinkcore}, Kitsak et al.\ propose a method
based on the notion of $k$-core to find the molecular complexes in
protein interaction networks. Altaf-Ul-Amin et al.\
\cite{03proteinfunkcore} propose a technique for predicting the
protein function based on $k$-core decomposition. In
\cite{05proteinkcore}, Wuchty and Almaas apply the $k$-core
decomposition to identify the layer structure of the protein
interaction network. In addition, the $k$-core decomposition is
recently used to identify the influential spreaders in complex
network \cite{10naturephykcore}. In \cite{10naturephykcore}, Kitsak
et al.\ find that the nodes located in the high order core are more
likely to be a influential spreader. Another line of research is to
investigate the $k$-core percolation in a random graph
\cite{06prlkcore, 06kcorepercolation, 11kcorepercolation}. These
studies mainly focus on investigating the threshold phenomenon of
the existence of a $k$-core based on some specific random graph
models.

From an algorithmic point of view, Batagelj and Zaversnik propose an
$O(n + m)$ algorithm for $k$-core decomposition in general graphs
\cite{03omalgkcore}. Their algorithm recursively deletes the node
with the lowest degree and uses the bin-sort algorithm to maintain
the order of the nodes. However, this algorithm needs to randomly
access the graph, thus it could be inefficient for the disk-resident
graphs. To overcome this problem, in \cite{11icdekcorejames}, Cheng
et al.\ propose an efficient $k$-core decomposition algorithm for
the disk-resident graphs. Their algorithm works in a top-to-down
manner that calculates the $k$-cores from higher order to lower
order. To make the $k$-core decomposition more scalable, in
\cite{11podcdistributedcore}, Montresor et al.\ propose a
distributed algorithm for $k$-core decomposition by exploiting the
locality property of $k$-core. All the above mentioned algorithms
are focus on $k$-core decomposition in static graph except for
\cite{10kcoredynamic}. For the dynamic graph, in
\cite{10kcoredynamic}, Miorandi and Pellegrini apply the $O(n+m)$
algorithm given in \cite{03omalgkcore} to recompute the core number
of the nodes when the graph is updated, which is clearly
inefficient. In the present paper, we propose a more efficient core
maintenance algorithm in dynamic graphs. Our algorithm are quite
efficient, which is more than 100 times faster than the
re-computation based algorithm.

\vspace*{-0.2cm}
\section{Conclusions}
\label{sec:concl} In this paper, we propose an efficient algorithm
for maintaining the core number of nodes in dynamic graphs. For a
node $u$, we define a notion of induced core subgraph $G_u$, which
contains the nodes that are reachable from $u$ and have the same
core number as $u$. Given a graph $G$ and an edge $(u, v)$, we find
that only the core number of nodes in $G_u$ or $G_v$ or $G_{u \cup
v}$ may need to be updated after inserting/deleing the edge $(u,
v)$. Based on this, first, we introduce a coloring algorithm to
identify all of these nodes. Second, we devise a recoloring
algorithm to determine the nodes whose core numbers definitely need
to be updated. Finally, we update the core number of such nodes by a
linear algorithm. In addition, we develop two pruning strategies,
namely $X$-pruning and $Y$-pruning, to further accelerate the
algorithm. We evaluate our algorithm over 15 real-world and 5 large
synthetic datasets. The results demonstrate the efficiency of our
algorithm.

\vspace*{-0.2cm} {
\bibliographystyle{abbrv}
\bibliography{kcore}
}

\end{document}